\documentclass[a4paper]{arxiv}

\usepackage{hyperref}
\usepackage[latin1]{inputenc}
\usepackage{xspace}
\usepackage{latexsym}
\usepackage{amsmath,amstext,amsxtra,amsgen,amsbsy,amsopn,amscd}
\usepackage{latexsym}
\usepackage{mathrsfs}
\usepackage{dsfont}
\usepackage[active]{srcltx}

\numberwithin{equation}{section}
\newtheorem{assumption}{Assumption}[section]

\def\Xint#1{\mathchoice
   {\XXint\displaystyle\textstyle{#1}}%
   {\XXint\textstyle\scriptstyle{#1}}%
   {\XXint\scriptstyle\scriptscriptstyle{#1}}%
   {\XXint\scriptscriptstyle\scriptscriptstyle{#1}}%
   \!\int}
\def\XXint#1#2#3{{\setbox0=\hbox{$#1{#2#3}{\int}$}
     \vcenter{\hbox{$#2#3$}}\kern-.5\wd0}}

\def\fint{\Xint-}

\newcommand{\ii}{\infty}
\newcommand{\Z}{\mathbb{Z}}
\newcommand{\R}{\mathbb{R}}
\newcommand{\N}{\mathbb{N}}

\newcommand{\CC}{\mathscr{C}}

\newcommand{\E}{\mathcal{E}}
\newcommand{\cE}{\mathcal{E}}
\newcommand{\cF}{\mathcal{F}}
\newcommand{\cL}{\mathscr{L}}

\newcommand{\LL}{\mathcal{L}}
\newcommand{\Q}{\mathcal{Q}}

\newcommand{\Ll}{\mathcal{L}}
\newcommand{\K}{\mathcal{K}}
\newcommand{\coul}{\mathcal{C}}

\newcommand{\ep}{\varepsilon}

\newcommand{\dd}{\partial}
\newcommand{\lam}{\lambda}
\newcommand{\supp}{\mathrm{supp}}
\DeclareMathOperator{\Tr}{{\rm Tr}}
\DeclareMathOperator{\Tro}{{\rm Tr}_0}
\newcommand{\norm}[1]{ \left| \! \left| #1 \right| \! \right| }

\newcommand{\tr}{\Tr}
\newcommand{\bral}{\left\langle}
\newcommand{\brar}{\right|}
\newcommand{\ketl}{\left|}
\newcommand{\ketr}{\right\rangle}
\newcommand{\ptgf}{\mathcal{E}^{\rm P}_{\dem}}
\newcommand{\ptge}{E^{\rm P}_{\dem}}
\newcommand{\ptgm}{\psi^{\rm P}_{\dem}}
\newcommand{\ptgint}{F^{\rm P}_{\dem}}

\newcommand{\FSm}{\gamma_{\rm per} ^0}
\newcommand{\FSh}{H_{\rm per} ^0}
\newcommand{\FSl}{\varepsilon_{\rm F}}
\newcommand{\FSd}{\rho_{\rm per} ^0}
\newcommand{\FSn}{\mu_{\rm per} ^0}
\newcommand{\FSp}{V_{\rm per} ^0}
\newcommand{\dem}{\varepsilon_{\rm M}}
\newcommand{\Wrho}{W_{\rho}}
\newcommand{\rhoQ}{\rho_Q}
\newcommand{\rhoP}{\rho_{\Psi}}
\newcommand{\trhoP}{\tilde{\rho}_{\Psi}}
\newcommand{\polf}{\mathcal{E}_m ^{\mathrm{pol}}}
\newcommand{\totf}{\mathcal{E}_m}
\newcommand{\tote}{E_m}

\newcommand{\crysint}{F_{\rm crys} }
\newcommand{\cryse}{\mathcal{F}_{\rm crys} }
\newcommand{\htotf}{\tilde{\mathcal{E}} _{m}}

\newcommand{\auxint}{F _{\mathrm{aux}}}
\newcommand{\Eperf}{\mathcal{E}^{\rm per}_m}
\newcommand{\Epere}{E ^{\rm per}_m}
\newcommand{\Eperm}{u_m^{\rm per}}
\newcommand{\fper}{f^{\rm per}}
\newcommand{\Eperelim}{E ^{\rm per}}
\newcommand{\Qpp}{Q ^{++}}
\newcommand{\Qpm}{Q ^{+-}}
\newcommand{\Qmp}{Q^{-+}}
\newcommand{\Qmm}{Q^{--}}
\newcommand{\one}{{\ensuremath {\mathds 1} }}
\newcommand{\oneep}{\one_{\left(-\infty, \FSl \right)}}
\newcommand{\oneepp}{\one_{\left( \FSl, +\infty \right)}}
\newcommand{\ointC}{\frac{1}{2i \pi}\oint_{\CC}}
\newcommand{\Ran}{\rm Ran}
\newcommand{\Ker}{\rm Ker}

\newcommand{\fintG}{\fint_{\Gamma ^*}}

\newcommand{\rhff}{\E ^{\rm rHF}}
\newcommand{\polM}{\Psi ^{\rm pol}_m}
\newcommand{\tpolM}{\tilde{\Psi} ^{\rm pol}_m}
\newcommand{\Sch}{\mathfrak{S}}
\newcommand{\FP}{\Psi^{\rm P}}
\newcommand{\EperM}{U ^{\rm per}_m}
\newcommand{\Ftest}{\Psi ^{\rm trial}}
\title{\Large Derivation of Pekar's Polarons from\\ \vspace{0.4cm} a Microscopic Model of Quantum Crystals}
\runningtitle{Derivation of Pekar's Polarons}
\author{Mathieu LEWIN}
\address{CNRS \& Department of Mathematics (UMR 8088), University of Cergy-Pontoise,\\ 95 000 Cergy-Pontoise,
France\\ Email: \email{mathieu.lewin@math.cnrs.fr}}
\author{Nicolas ROUGERIE}
\address{CNRS \& Department of Mathematics (UMR 8088), University of Cergy-Pontoise,\\ 95 000 Cergy-Pontoise,
France\\ Email: \email{nicolas.rougerie@u-cergy.fr}}

\date{August 30, 2011}

\begin{document}

\maketitle

\bigskip

\begin{abstract}
A polaron is an electron interacting with a polar crystal, which is able to form a bound state by using the distortions of the crystal induced by its own density of charge. In this paper we derive Pekar's famous continuous model for polarons (in which the crystal is replaced by a simple effective Coulomb self-attraction) by studying the macroscopic limit of the reduced Hartree-Fock theory of the crystal. The macroscopic density of the polaron converges to that of Pekar's nonlinear model, with a possibly anisotropic dielectric matrix. The polaron also exhibits fast microscopic oscillations which contribute to the energy at the same order, but whose characteristic length is small compared to the scale of the polaron. These oscillations are described by a simple periodic eigenvalue equation. Our approach also covers multi-polarons composed of several electrons, repelling each other by Coulomb forces.

\medskip

\noindent{\scriptsize\copyright~2011 by the authors. This paper may be reproduced, in its entirety, for non-commercial~purposes.}
\end{abstract}

\tableofcontents

\bigskip

%%%%%%%%%%%%%%%%%%%%%%%%%%%%%%%%%%%%%%%%%%%%%%%%%%%%%%
%%%%%%%%%%%%%%%%%%%%%%%%%%%%%%%%%%%%%%%%%%%%%%%%%%%%%%
\section*{Introduction}
\addcontentsline{toc}{section}{Introduction}
%%%%%%%%%%%%%%%%%%%%%%%%%%%%%%%%%%%%%%%%%%%%%%%%%%%%%%
%%%%%%%%%%%%%%%%%%%%%%%%%%%%%%%%%%%%%%%%%%%%%%%%%%%%%%

In vacuum, $N$ electrons cannot form a bound state, because of their Coulomb repulsion and the dispersive nature of their kinetic energy. Confining the electrons in a given region of space is only possible by applying an external field. In atoms and molecules, this field is the electrostatic attraction of the (positively charged) nuclei. 
The situation is completely different when the electrons are placed in a polarizable medium like a dielectric crystal. There, the $N$ electrons induce a lattice distortion by repelling a bit the negative ions of the crystal and attracting the positive ones.
When the resulting polarization is strong enough, the electrons are able to overcome their Coulomb repulsion and to form a bound state, called an \emph{$N$-polaron}.

Polarons have been widely studied in the physics literature. The main difficulty is to adequately describe the behavior of the polar crystal and of its distortions. The simplest model was obtained by Pekar~\cite{Pekar-54,Pekar-63} who proposed to approximate the crystal by a continuous polarizable medium, described only by its static and high frequency dielectric constants. For the polaron, composed of only one electron, one gets the energy functional~\cite{AleDev-09,BecNieRuiSol-00}
\begin{equation}
\cE^{\rm P}_{\varepsilon_{\rm M}}[\psi,D] = \frac{1}{2}\int_{\R^3}|\nabla\psi(x)|^2\,dx -  \int_{\R^3}|\psi(x)|^2v(x)\,dx + \frac{1}{8\pi\big((\dem)^{-1}-1\big)}\int_{\R^3}|D(x)|^2dx.
\label{eq:polaron-in-polarization-field}
\end{equation}
Here $\psi$ is the wave function of the electron, $D=-\nabla v$ is the displacement field of the continuous medium, $\varepsilon_{\rm M}$ is its static dielectric constant, in units such that the high frequency dielectric constant is $\varepsilon_0=1$ and such that the charge and the mass of the electron are normalized to $e=1$ and $m_{e^-}=\hbar^2$. For simplicity we neglect the spin of the electron in the whole paper. Minimizing the above functional with respect to $D$ at fixed $\psi$, we obtain the following energy functional for the polaron alone in the continuous polarizable medium
\begin{equation}
\cE^{\rm P}_{\varepsilon_{\rm M}}[\psi]=\frac{1}{2}\int_{\R^3}|\nabla\psi(x)|^2\,dx+\frac{(\varepsilon_{\rm M})^{-1}-1}{2}\int_{\R^3}\int_{\R^3}\frac{|\psi(x)|^2|\psi(y)|^2}{|x-y|}dx\,dy.
\label{eq:Pekar-intro}
\end{equation}
The last nonlinear term in~\eqref{eq:Pekar-intro} is an effective Coulomb self-interaction. When $\varepsilon_{\rm M}>1$, this nonlinear term is attractive and it has been shown by Lieb~\cite{Lieb-77} that the Pekar energy functional~\eqref{eq:Pekar-intro} admits a unique minimizer up to translations (under the normalization constraint $\int_{\R^3}|\psi|^2=1$). This ground state is radial and solves the corresponding Euler-Lagrange equation
\begin{equation}
\left(-\frac{\Delta}{2}+\big((\varepsilon_{\rm M})^{-1}-1\big)|\psi|^2\star|x|^{-1}\right)\psi=E\,\psi.
\label{eq:Pekar}
\end{equation}
This nonlinear equation is ubiquitous in Physics and it is sometimes also called the ``Choquard'' or ``Schr\"odinger-Newton'' equation.

Pekar's theory can be easily generalized to the case of $N$ electrons, as was first suggested for $N=2$ by Pekar and Tomasevich~\cite{PekTom-51}. Taking into account the electrostatic repulsion between the electrons, one arrives at the following many-body energy functional: 
\begin{multline}
\cE^{\rm P}_{\varepsilon_{\rm M}}[\Psi]=\int_{\R^{3N}}\left(\frac12\sum_{j=1}^N \left|\nabla_{x_j}\Psi(x_1,...,x_N)\right|^2+\sum_{1\leq k<\ell\leq N}\frac{|\Psi(x_1,...,x_N)|^2}{|x_k-x_\ell|}\right)dx_1\cdots dx_N\\
+\frac{(\varepsilon_{\rm M})^{-1}-1}{2}\int_{\R^3}\int_{\R^3}\frac{\rho_\Psi(x)\rho_\Psi(y)}{|x-y|}dx\,dy.
\label{eq:Pekar-N-intro}
\end{multline}
The many-body wave function $\Psi$ must be antisymmetric with respect to exchanges of the variables $x_1,...,x_N$, due to the fermionic nature of the electrons. Also, $\rho_\Psi$ is the total density of the $N$ electrons, defined by
\begin{equation}\label{defi densite pre}
\rho_\Psi(x)=N\int_{\R^3}dx_2\cdots\int_{\R^3}dx_N\;|\Psi(x,x_2,...,x_N)|^2.
\end{equation}
There is now a competition between the many-body electronic repulsion and the nonlinear attraction due to the polarizable medium. It has been shown recently by one of us~\cite{Lewin-11} that when $\varepsilon_{\rm M}$ is sufficiently large (depending on $N$), the many-body Pekar functional~\eqref{eq:Pekar-N-intro} admits at least one minimizer, hence infinitely many by translation invariance. On the other hand, it can be deduced from the results of Frank, Lieb, Seiringer and Thomas~\cite{FraLieSeiTho-10a,FraLieSeiTho-10b} that when $\varepsilon_{\rm M}\leq 1+a$ (with $a$ independent of $N$), $\cE^{\rm P}_{\varepsilon_{\rm M}}$ has no ground state for $N\geq2$.

Pekar's functional is not the only one used by physicists to describe ($N$-)polarons interacting with a continuous medium. In~\cite{Frohlich-37,Frohlich-52}, H. Fröhlich has proposed to replace the classical polarization field $D$ of~\eqref{eq:polaron-in-polarization-field}, by a quantized (phonon) field with which the electrons interact~\cite{AleDev-09}. Fröhlich's model has been mathematically studied in several works. In particular, Donsker and Varadhan ~\cite{DonVar-83} and then, with a different approach, Lieb and Thomas \cite{LieTho-97}, have proved that Pekar's polaron can be recovered from the strong coupling limit of Fröhlich's model. This was later extended to bi-polarons by Miyao and Spohn in~\cite{MiySpo-07}. For other recent works on Fröhlich's and Pekar's theories, see for instance~\cite{Moller-06,GriMol-10,FraLieSeiTho-10a,FraLieSeiTho-10b,BenBle-11}.

Both models assume that the medium in which the particle evolve is continuous. In a crystal, this can only be valid when the size of the electronic system is much bigger than the typical lattice length, that is, the diameter of the unit cell. One then speaks of \emph{large polarons}. For smaller polarons, this approximation is not good enough and the electrons start to see the detailed structure of the crystal.

\medskip

The purpose of this paper is twofold. First, based on previous works by the first author with Cancès and Deleurence~\cite{CanDelLew-08a,CanDelLew-08b,CanLew-10}, we write a simple mean-field model describing small multi-polarons in a quantum crystal. Second, we show that in a macroscopic limit where the system lives on a scale much larger than the size of the lattice cell, we again recover the Pekar(-Tomasevich) theory. We now explain the main lines of our approach, before turning to a more detailed presentation of our results in the next section.

Our crystal is assumed to be extended over the whole space. At rest, it is composed of classical nuclei, described by an $\cL$-periodic charge density $\mu^0_{\rm per}\geq0$, and of quantum electrons which are modelled by an $\cL$-periodic density $\rho^0_{\rm per}$. The lattice $\cL$ is a discrete subgroup of $\R^3$ whose fundamental domain $\Gamma$ (the unit cell) is compact, for instance $\cL=\Z^3$. The system is locally neutral in the sense that $\int_{\Gamma}\rho^0_{\rm per}=\int_{\Gamma}\mu^0_{\rm per}$. The unperturbed crystal induces an $\cL$-periodic electrostatic potential $V^0_{\rm per}$ which solves Poisson's equation
$$-\Delta V^0_{\rm per}=4\pi\big(\rho^0_{\rm per}-\mu^0_{\rm per}\big).$$
This electrostatic potential is felt by any other particle which is added to the system. 

When the additional particles are inserted, the nuclei and the electrons of the crystal can be displaced a little bit. This distortion is described by (local) perturbations $\delta\mu$ and $\delta\rho$ such that the nuclear and electronic densities become $\mu=\mu_{\rm per}^0+\delta\mu$ and $\rho=\rho^0_{\rm per}+\delta\rho$. 
The inserted particles then feel the electrostatic field $(\delta\rho-\delta\mu)\star|x|^{-1}$ induced by these displacements.
For the single polaron, we are thus led to an energy functional of the form
\begin{multline}
\cE[\psi,\delta\rho,\delta\mu]=\frac1{2}\int_{\R^3}|\nabla\psi(x)|^2\,dx+\int_{\R^3}V^0_{\rm per}(x)|\psi(x)|^2\,dx\\+\int_{\R^3}\int_{\R^3}\frac{\big(\delta\rho(x)-\delta\mu(x)\big)|\psi(y)|^2}{|x-y|}\,dx\,dy+\cF_{\rm crys}[\delta\rho,\delta\mu]
\label{eq:abstract-energy}
\end{multline}
where $\cF_{\rm crys}[\delta\rho,\delta\mu]$ is the energy cost to perturb the crystal by moving the nuclei of $\delta\mu$ and the electrons of $\delta\rho$. Note that the periodic density $\rho^0_{\rm per}-\mu^0_{\rm per}$ is locally neutral, and that the displacement densities $\delta\rho$ and $\delta\mu$ should satisfy 
$$\int_{\R^3}\delta\rho(x)\,dx=\int_{\R^3}\delta\mu(x)\,dx=0,$$
at least formally (see Remark \ref{rmk:epsilon_F} below). As expected, the polaron effectively sees an electrostatic potential induced by a field of dipoles.

Our two subsystems (the polaron and the crystal) are uncorrelated in this simplified theory, hence it is possible to completely eliminate the crystalline degrees of freedom, by minimizing over $\delta\mu$ and $\delta\rho$ for any fixed state $\psi$ of the polaron. This leads to an effective nonlinear functional for the electron alone, of the form
\begin{equation}
\cE_{\rm eff}[\psi]=\frac1{2}\int_{\R^3}|\nabla\psi(x)|^2\,dx+\int_{\R^3}V^0_{\rm per}(x)|\psi(x)|^2\,dx+F_{\rm crys}\big[|\psi|^2\big],
\label{eq:Pekar-abstract-intro}
\end{equation}
where the nonlinear effective energy $F_{\rm crys}$ is defined by
\begin{equation}
 F_{\rm crys}\big[|\psi|^2\big]=\inf_{\substack{\delta\rho\geq-\rho^0_{\rm per}\\ \delta\mu\geq-\mu^0_{\rm per}}}\left(\int_{\R^3}\int_{\R^3}\frac{\big(\delta\rho(x)-\delta\mu(x)\big)|\psi(y)|^2}{|x-y|}\,dx\,dy+\cF_{\rm crys}[\delta\rho,\delta\mu]\right).
\label{eq:def_nonlinearity_intro}
\end{equation}
The case of the $N$-polaron is obviously similar, leading to the effective nonlinear many-body functional
\begin{multline}
\cE_{\rm eff}[\Psi]=\int_{\R^{3N}}\left(\frac1{2}\sum_{j=1}^N|\nabla_{x_j}\Psi(x_1,...,x_N)|^2+\sum_{1\leq k<\ell\leq N}\frac{|\Psi(x_1,...,x_N)|^2}{|x_k-x_\ell|}\right)dx_1\cdots dx_N\\
+\int_{\R^3}V^0_{\rm per}(x)\rho_{\Psi}(x)\,dx+F_{\rm crys}\big[\rho_{\Psi}\big],
\label{eq:Pekar-N-abstract-intro}
\end{multline}
where $F_{\rm crys}$ is the same nonlinearity as in~\eqref{eq:def_nonlinearity_intro}.

Of course, the main difficulty in this context is to find an energy functional $\cF_{\rm crys}[\delta\rho,\delta\mu]$ quantifying the cost to move nuclei and electrons in the crystal, which is both physically relevant and mathematically amenable.
It turns out that there is a well-defined mean-field theory~\cite{CanDelLew-08a} for the electronic perturbation $\delta\rho$ for every fixed value of $\delta\mu$, but that allowing the nuclei to move freely is too involved for the present mathematical technology. If the crystal is not globally stable with respect to the positions of the nuclei, the latter will want to relax to better positions, changing thereby the total energy per unit volume and rendering our above effective energy $F_{\rm crys}$ infinite. 

To face this problem we could add some stability conditions on the nuclear structure of our crystal, but this would complicate our exposition dramatically. Since in this paper we are more interested in the derivation of Pekar's polaron from a microscopic model than in proposing a quantitative theory, we will make the (very strong) assumption that \emph{the nuclei cannot move}, $\delta\mu\equiv0$. As we will see, the distortion of the electronic Fermi sea of the crystal is in principle enough to bind polarons, although its effect is weaker than when the nuclear displacements are taken into account. As far as the derivation of Pekar's polarons is concerned, this simply means that we will only obtain the electronic contribution to the dielectric constant $\varepsilon_{\rm M}$. 

In Section~\ref{sec:model-crystal} below, following~\cite{CanDelLew-08a}, we define an appropriate functional $F_{\rm crys}$ obtained in a Hartree-Fock-type approximation, when only the electronic Fermi sea is allowed to move. In this introduction we assume that $F_{\rm crys}$ is given to us without giving its precise expression, and we now discuss the derivation of Pekar's polarons in a macroscopic limit. Of course the precise form of $F_{\rm crys}$ is very important, since it is at the origin of the dielectric constant $\varepsilon_{\rm M}$ seen in Pekar's theory.

We now explain our derivation of Pekar's energy functional in the case of only one electron (polaron), the argument being similar for the $N$-polaron. We have to let the polaron live on a much larger scale than the typical size of the lattice cell or, equivalently, to make the crystal live on a much smaller scale than that of the polaron. To this end, we introduce a small parameter $0<m\ll1$ which is interpreted as the ratio between the microscopic and the macroscopic lengths. At the macroscopic scale, the lattice becomes $m\cL$ and the energy functional of the polaron is now
\begin{equation}
\cE_{m}[\psi]=\frac1{2}\int_{\R^3}|\nabla\psi(x)|^2\,dx+m^{-1}\int_{\R^3}V^0_{\rm per}(x/m)|\psi(x)|^2\,dx+m^{-1}F_{\rm crys}\big[m^{3}|\psi(m\cdot)|^2\big].
\label{eq:Pekar-abstract-intro-macro}
\end{equation}
The scaling is chosen to make all the terms of the energy contribute the same at the macroscopic level. As for the periodic potential of the crystal, we have that $V_m:=m^{-1}V^0_{\rm per}(x/m)$ is the unique solution of Poisson's equation
$$-\Delta V_m=m^{-3}\left(\rho^0_{\rm per}(x/m)-\mu^0_{\rm per}(x/m)\right).$$
The scaling of the density is chosen such as to keep constant the number of electrons and nuclei per unit cell.\footnote{That is, we have $\int_{m\Gamma}m^{-3}\rho^0_{\rm per}(\cdot/m)=\int_\Gamma\rho^0_{\rm per}$ and $\int_{m\Gamma}m^{-3}\mu^0_{\rm per}(\cdot/m)=\int_\Gamma\mu^0_{\rm per}$.}
There are similar arguments in favour of the chosen scaling $m^{-1}F_{\rm crys}\big(m^{3}|\psi(m\cdot)|^2\big)$ for the nonlinear term.

Changing variables $\tilde\psi=m^{3/2}\psi(m\cdot)$ we can express the same functional at the microscopic scale
\begin{equation}
\cE_{m}[\tilde\psi]=m^{-1}\left(\frac1{2m}\int_{\R^3}|\nabla\tilde\psi(x)|^2\,dx+\int_{\R^3}V^0_{\rm per}(x)|\tilde\psi(x)|^2\,dx+F_{\rm crys}\big[|\tilde\psi|^2\big]\right).
\label{eq:Pekar-abstract-intro-micro}
\end{equation}
From the perspective of the crystal, a large polaron can therefore be obtained by inserting a particle whose mass $m$ is very small, and which thus tends to be very spread out in space.

For an isotropic crystal described by the nonlinear energy $F_{\rm crys}$ defined later, we prove in Theorem~\ref{theo:PTg1} below that in the limit $m\to0$, any ground state of~\eqref{eq:Pekar-abstract-intro-macro} behaves as follows
\begin{equation}
\boxed{\phantom{\int}\psi_m(x)\underset{m\to0}{\simeq} u_m^{\rm per}(x/m)\; \psi^{\rm P}_{\varepsilon_{\rm M}}(x),\phantom{\int}}
\label{eq:limit_fn_intro}
\end{equation}
up to a well-chosen translation of the system in space.
Here $\psi^{\rm P}_{\varepsilon_{\rm M}}$ is the unique ground state of Pekar's functional~\eqref{eq:Pekar-intro}, $\varepsilon_{\rm M}>1$ being the macroscopic dielectric constant of the crystal, which will be defined in Section~\ref{sec:model-crystal}. 
On the other hand, $u_m^{\rm per}$ is an $\cL$-periodic function, which converges to $1$ uniformly as $m\to0$. It is defined by minimizing the functional
\[
\Eperf [v] = \int_{\Gamma} \frac{1}{2m}|\nabla v| ^2 + \FSp |v| ^2
\]
with periodic boundary conditions on $\dd \Gamma$ and  under the constraint that $\int_{\Gamma} |\Eperm| ^2 = \left|\Gamma \right|$, where we recall that $\Gamma$ is the unit cell of the lattice. Extended by periodicity over the whole of $\R ^3$ it solves
\begin{equation}\label{eq:equation_u_per}
\left(-\frac{\Delta}{2m}+V^0_{\rm per}(x)\right)u_m^{\rm per}(x)=  \Epere\, u_m^{\rm per}(x).
\end{equation}
The precise behavior of $u_m^{\rm per}$ as $m\to0$ (as well as the value of $\Epere$) can be determined by usual perturbation theory, as we will explain later in Section~\ref{sec:periodic}. 
The corresponding energy of $\psi_m$ tends to the sum of the energies of the two functions $u_m^{\rm per}$ and $\psi^{\rm P}_{\varepsilon_{\rm M}}$:
\begin{equation}
\lim_{m\to0}\cE_{m}[\psi_m]=E^{\rm per}+\cE^{\rm P}_{\varepsilon_{\rm M}}\left[\psi^{\rm P}_{\varepsilon_{\rm M}}\right]
\label{eq:limit_energy_intro}
\end{equation}
where $\Eperelim$ is the limit of $m ^{-1} \Epere$ when $m\to 0$ and $\ptgf$ is Pekar's energy~\eqref{eq:Pekar-intro}.

Let us emphasize that the limit studied in this paper is completely different from existing results on Fröhlich's model~\cite{DonVar-83,LieTho-97,MiySpo-07}. In Fröhlich's theory, $\dem$ is a parameter to be chosen. The goal is to show that in the strong coupling limit the polaron tends to decouple from the quantized field, leading to Pekar's ground state with the given $\dem$. In this paper the correlations between the polaron and the crystal are already neglected and our purpose is to derive Pekar's model in a macroscopic limit. Our derivation provides a certain value for the dielectric constant $\dem$, in terms of the structure of the chosen microscopic quantum crystal.

In this introduction we have explained the simplest situation of an isotropic crystal whose dielectric tensor $\varepsilon_{\rm M}$ is a constant. Below we consider the general case and, in the anisotropic case, we obtain a generalized Pekar functional in which $\varepsilon_{\rm M}$ is a $3\times3$ real symmetric matrix. The corresponding expression for $\cE^{\rm P}_{\dem}$ will be given in Section~\ref{sec:generalized-Pekar} below. Also, the results discussed here for the single polaron hold similarly for $N$-polarons.

\medskip

The paper is organized as follows. In Section \ref{sec:main-results} we describe our small polaron model in detail. We then recall some facts about Pekar's theory and state our main results relating the latter to the reduced Hartree-Fock theory of quantum crystals. Their proofs rely on two main ingredients. First, in Section \ref{sec:periodic} we separate out the contribution of the microscopic oscillations, by using a simple energy decoupling argument and the properties of the periodic eigenvalue problem. The second, more involved, step is the detailed analysis of the limit of the perturbed crystal model of \cite{CanDelLew-08a}. Section \ref{sec:effective-interaction} proceeds with improving some results of \cite{CanLew-10} and applying them to the context of the polarons. Finally, we complete the proofs of our main results in Section \ref{sec:completion}.

\bigskip

\noindent\textbf{Acknowledgment.} The research leading to these results has received funding from the European Research Council under the European Community's Seventh Framework Programme (FP7/2007-2013 Grant Agreement MNIQS No. 258023).

%%%%%%%%%%%%%%%%%%%%%%%%%%%%%%%%%%%%%%%%%%%%%%%%%%%%%%
%%%%%%%%%%%%%%%%%%%%%%%%%%%%%%%%%%%%%%%%%%%%%%%%%%%%%%
\section{Main results}\label{sec:main-results}

In this section we introduce some preliminary tools which are necessary to properly define our model, and we state our main theorems. Proofs will be given in the next sections.

%%%%%%%%%%%%%%%%%%%%%%%%%%%%%%%%%%%%%%%%%%%%%%%%%%%%%%
%%%%%%%%%%%%%%%%%%%%%%%%%%%%%%%%%%%%%%%%%%%%%%%%%%%%%%
\subsection{Energy to perturb the interacting periodic Fermi sea}\label{sec:model-crystal}

In this section we quickly present the model that Cancès, Deleurence and Lewin have introduced in~\cite{CanDelLew-08a,CanDelLew-08b} to describe the distortions of the electrons in a quantum crystal, using a Hartree-Fock-type theory. We will define the effective nonlinear energy $F_{\rm crys}$ felt by the polaron, and also quickly recall the definition of the macroscopic dielectric tensor $\varepsilon_{\rm M}$ which will later appear in Pekar's functional. We provide some additional technical details later in Section~\ref{sec:input}.

We fix an $\cL$-periodic density of charge $\mu^0_{\rm per}$ for the classical nuclei of the crystal, with $\cL$ a discrete subgroup of $\R^3$. It is enough for our purpose to assume that $\mu^0_{\rm per}$ is a locally-finite non-negative measure, such that
$\int_{\Gamma}\mu_{\rm per}^0=Z\in\N$, where $\Gamma=\R^3/\cL$ is the unit cell.

In reduced Hartree-Fock theory~\cite{Solovej-91}, the state of the electrons in the crystal is described by a \emph{one-particle density matrix}, which is a self-adjoint operator $\gamma:L^2(\R^3)\to L^2(\R^3)$ such that $0\leq\gamma\leq1$ (in the sense of operators). When no external field is applied to the system, the electrons arrange in a periodic configuration $\gamma=\gamma^0_{\rm per}$, which is a solution of the \emph{reduced Hartree-Fock equations}
\begin{equation}
\begin{cases}
\gamma^0_{\rm per}=\oneep\big(-\Delta/2+V^0_{\rm per}\big),\\[0.2cm]
-\Delta V^0_{\rm per}=4\pi\big(\rho_{\gamma^0_{\rm per}}-\mu^0_{\rm per}\big),\\[0.2cm]
\displaystyle\int_{\Gamma}\rho _{\gamma^0_{\rm per}}=\int_\Gamma\mu^0_{\rm per}.
\end{cases}
\label{eq:rHF-periodic}
\end{equation}
Here $\rho_{A}$ denotes the density of the operator $A$ which is formally given by 
$\rho_A(x)=A(x,x)$ when $A$ is locally trace-class. Also, $\oneep(H)$ denotes the spectral projector of $H$ onto the interval $(-\ii,\FSl)$. The real number $\FSl$ in~\eqref{eq:rHF-periodic} is called the \emph{Fermi level}. It is also the Lagrange multiplier used to impose the constraint that the system must be locally neutral (third equation in~\eqref{eq:rHF-periodic}). It should not be confused with the dielectric constant $\dem$ which will be defined later. 
Existence and uniqueness of solutions to the self-consistent equation~\eqref{eq:rHF-periodic} was proved in~\cite{CatBriLio-01,CanDelLew-08a} (see also~\cite{Nier-93} at positive temperature).

By the Bloch-Floquet theory~\cite{ReeSim1}, the spectrum of the $\cL$-periodic Schrödinger operator 
$$H^0_{\rm per}=-\frac12 \Delta+V^0_{\rm per}(x)$$
is composed of bands. When there is a gap between the $Z$th and the $(Z+1)$st bands, the crystal is an insulator and $\FSl$ can be any arbitrary number in the gap. Like in~\cite{CanDelLew-08a}, in the whole paper we will assume that the host crystal is an insulator.

\begin{assumption}[\textbf{The host crystal is an insulator}]\label{asum:insulator}\ \\
\it
The periodic Schrödinger operator $H^0_{\rm per}$ has a gap between its $Z$th and $(Z+1)$st bands, and we fix any chemical potential $\FSl$ in the corresponding gap.
\end{assumption}

When the quantum crystal is submitted to an external field, the Fermi sea polarizes. The new density matrix $\gamma$ of the system now solves the nonlinear equation
\begin{equation}
\gamma=\oneep\Big(-\Delta/2+V^0_{\rm per}+(\rho_{\gamma-\gamma^0_{\rm per}}-\nu)\star|x|^{-1}\Big)+\delta,
\label{eq:rHF-defect}
\end{equation}
where $\nu$ denotes the external density used to perturb the Fermi sea, and where the operator $\delta$ (satisfying $0\leq\delta\leq1$) lives only at the Fermi level\footnote{This means $\Ran(\delta) \subset \Ker \big(-\Delta/2+V^0_{\rm per}+(\rho_{\gamma-\gamma^0_{\rm per}}-\nu)\star|x|^{-1}-\FSl \big)$. The operator $\delta$ can safely be ignored by the reader, as in the macroscopic limit that we will later consider, we will always have $\delta \equiv 0$.}. Existence of solutions to this equation was shown in~\cite{CanDelLew-08a}. In general there is no uniqueness of $\gamma$, but the density $\rho_{\gamma}$ is itself unique. The perturbed state $\gamma$ is such that $Q:=\gamma-\gamma^0_{\rm per}$ is Hilbert-Schmidt and locally trace-class, and $\rho_Q\in L^2(\R^3)$. In general, $Q$ is \emph{not} even a trace-class operator~\cite{CanLew-10}. This motivates the introduction of a particular functional setting, some of its elements being recalled below (see also Section~\ref{sec:input}).

The method used in~\cite{CanDelLew-08a} to construct solutions is variational and it relies on an energy functional which we want to use in our polaron model. The idea is to define the energy cost to move the electrons from $\gamma^0_{\rm per}$ to $\gamma$ as the (formal) difference between the (infinite) reduced Hartree-Fock energies of $\gamma$ and of $\gamma^0_{\rm per}$. Denoting by
\begin{equation}
 \label{Coulomb}
D(f,g) := \iint_{\R^3 \times \R ^3} \frac{f (x)f (y)}{|x-y|} dxdy = 4\pi\int _{\R^3} \frac{\overline{\hat{f} (k)} \hat{g} (k)}{|k| ^2} \, dk
\end{equation}
the Coulomb interaction, the formal calculation is
\begin{multline}
\rhff [\gamma] - \rhff [\FSm]\text{``}=\text{''}
\;\left(\frac12\tr(-\Delta-\FSl)\gamma+\frac12 D(\rho_\gamma-\mu^0_{\rm per},\rho_\gamma-\mu^0_{\rm per})\right)\\ - \left(\frac12\tr(-\Delta-\FSl)\gamma^0_{\rm per}+\frac12 D(\rho_{\gamma^0_{\rm per}}-\mu^0_{\rm per},\rho_{\gamma^0_{\rm per}}-\mu^0_{\rm per})\right)\\
\text{``}=\text{''} \tr (H^0_{\rm per}-\FSl)Q + \frac12 D(\rho_Q,\rho_Q)
\label{eq:formal-calculation}
\end{multline}
with again $Q=\gamma-\gamma^0_{\rm per}$. 
Note that $Q$ satisfies $-\gamma^0_{\rm per}\leq Q\leq 1-\gamma^0_{\rm per}$, which is equivalent to
$Q^2\leq Q^{++}-Q^{--}$
with $Q^{++}:=(\gamma^0_{\rm per})^\perp Q (\gamma^0_{\rm per})^\perp\geq0$ and $Q^{--}:=\gamma^0_{\rm per} Q \gamma^0_{\rm per}\leq0$.
This allows to properly define the kinetic energy in~\eqref{eq:formal-calculation} as follows~\cite{CanDelLew-08a}: 
\begin{equation}\label{eq:gen kin ener}
\tr_0 (H^0_{\rm per}-\FSl)Q=\tr|H^0_{\rm per}-\FSl|^{1/2}\big(Q^{++}-Q^{--}\big)|H^0_{\rm per}-\FSl|^{1/2}.
\end{equation}
More generally, one can define the \emph{generalized trace} as 
\begin{equation}\label{eq:gen trace}
\Tro Q = \Tr \Qpp + \Tr \Qmm
\end{equation}
when $\Qpp$ and $\Qmm$ are trace-class. 
 
The relative energy \eqref{eq:gen kin ener} is $\geq0$ for every $\FSl$ in the band gap. The total energy to go from $\gamma^0_{\rm per}$ to $\gamma$ is defined as
\begin{equation}
\cF_{\rm crys}[Q]:=\tr_0 (H^0_{\rm per}-\FSl)Q+\frac12 D(\rho_Q,\rho_Q)
\label{eq:def_energy_crystal} 
\end{equation}
and it is also non-negative since we have
$$D(\rho,\rho)=4\pi\int_{\R^3}\frac{|\widehat{\rho}(k)|^2}{|k|^2}\,dk\geq0.$$
Taking into account the term involving the chemical potential $\FSl$ in the total energy allows to have a non-negative energy functional (the reference periodic Fermi sea now has energy zero). The precise value of $\FSl$ in the gap does not matter at this stage.

When we submit the crystal to an external density $\nu$, the state of the Fermi sea is obtained by solving the following minimization problem\footnote{In the whole paper we use a sign convention for the external density $\nu$ which is opposite to that of~\cite{CanDelLew-08a,CanLew-10}. In these works $\nu$ was interpreted as a \emph{nuclear} defect density, whereas in our case it will be that induced by our polaron, which are negatively charged particles.}
\begin{equation}
\boxed{F_{\rm crys}[\nu]=\inf_{-\gamma^0_{\rm per}\leq Q\leq 1-\gamma^0_{\rm per}}\Big(D(\nu,\rho_Q)+\cF_{\rm crys}[Q]\Big).}
\label{eq:def_F_crys}
\end{equation}
As shown in~\cite{CanDelLew-08a}, for any $\nu\in L^1(\R^3)\cap L^{2}(\R^3)$, this minimization problem has at least one solution $Q=\gamma-\gamma^0_{\rm per}$ (in an adequate function space that we recall in Section \ref{sec:aux_interaction}), with $\gamma$ solving the nonlinear equation~\eqref{eq:rHF-defect}. The corresponding density $\rho_Q$ is in $L^2(\R^3)$ but in general it has long range oscillations which are not integrable at infinity~\cite{CanLew-10}. The total energy $F_{\rm crys}[\nu]$ of the crystal submitted to an external density $\nu$ is the one which we will use later, with $\nu=|\psi|^2$ (polaron) or $\nu=\rho_\Psi$ ($N$-polaron).

\begin{remark}\label{rmk:epsilon_F}
Instead of working in a grand canonical formalism where $\FSl$ has a fixed value, we could impose that no charge can be added to the crystal which, in this context, means $\tr_0(Q)=0$ (recall that $\rho_Q$ is not necessarily integrable at infinity). It has been shown in~\cite{CanLew-10} (Lemma 5) that when the external density $\nu$ is small enough, e.g. $D(\nu,\nu)\leq\eta$, then we have automatically $\tr_0(Q)=0$ for every $\FSl$ which is at a distance $\geq C\sqrt{\eta}$ of the edges of the gap. The value of $F_{\rm crys}[\nu]$ is independent of $\FSl$ in this range. 

In the macroscopic limit that we consider later, the polaron induces a charge defect in the crystal which, at the microscopic scale, is very spread out in space and we will automatically have that the Fermi sea stays neutral, $\tr_0(Q)=0$. As for the derivation of Pekar's polaron, the precise value of $\FSl$ in the gap therefore does not matter.
\end{remark}

\begin{remark}
Using well-known ideas from Density Functional Theory~\cite{Lieb-83b}, one can express the whole problem only in terms of the perturbed density $\delta\rho=\rho_Q$, as we have done in the introduction. The energy cost to perturb the Fermi sea by a density $\delta\rho$ is defined as
$$\cF^{\,'}_{\rm crys}[\delta\rho]=\inf_{\substack{-\gamma^0_{\rm per}\leq Q\leq 1-\gamma^0_{\rm per}\\ \rho_Q=\delta\rho}} \cF_{\rm crys}[Q],$$
such that $F_{\rm crys}[\nu]$ can also be expressed as
$$F_{\rm crys}[\nu]=\inf_{\delta\rho\geq-\rho^0_{\rm per}}\Big(D(\nu,\rho_Q)+\cF^{\,'}_{\rm crys}[\delta\rho]\Big).$$
\end{remark}

When the crystal is submitted to the external density $\nu$, the corresponding total energy $F_{\rm crys}[\nu]$ satisfies the following simple estimate.

\begin{lemma}[A uniform estimate on $F_{\rm crys}$]\label{lem:Fcrys}\mbox{}\\
We have for all $\nu\in L^1(\R^3)\cap L^{6/5}(\R^3)$
\begin{equation}
-\frac12 D(\nu,\nu)\leq F_{\rm crys}[\nu]\leq 0.
\label{eq:lower-bound-F-crys}
\end{equation}
\end{lemma}

\begin{proof}
The upper bound is obtained by taking $Q=0$ as test state in~\eqref{eq:def_F_crys}. For the lower bound we neglect the positive kinetic energy of $Q$,  complete the square and use that $D(\cdot,\cdot)$ defines a scalar product.
\end{proof}

The estimate~\eqref{eq:lower-bound-F-crys} already has interesting physical consequences for the polaron. The fact that $\crysint$ is non-positive means that the interaction of the electron with the crystal will always be attractive. On the other hand, the lower bound means that Pekar's interaction (with $\dem\equiv\ii$) is always a lower estimate to the total interaction with the crystal.

However, we will see below that~\eqref{eq:lower-bound-F-crys} is not optimal in the macroscopic limit, corresponding to density $\nu$'s which are very spread-out in space. The correct macroscopic behavior was studied by Cancès and Lewin in~\cite{CanLew-10}. 
In particular, they have derived the (electronic) dielectric matrix $\dem$ of the crystal, which is a $3\times3$ symmetric real matrix such that $\dem>1$ in the sense of matrices. 

The dielectric matrix $\dem$ can be expressed in terms of the Bloch transform of $H^0_{\rm per}$, see Eq. (36) in~\cite{CanLew-10}. 
The formula of $\dem$ is well known~\cite{Adler-62,Wiser-63} but it is not really important to us. 
More important is Theorem 3 of~\cite{CanLew-10}, which says that $\dem$ can be obtained by a macroscopic excitation of the Fermi sea: If we take an external density of the form 
\[
\nu_m(x)=m^3\,\nu(mx)                                                                                                                                                                                                                                                                                                                                                                 \]
and call $Q_m$ one corresponding solution of~\eqref{eq:rHF-defect}, then the rescaled self-consistent potential 
\begin{equation}\label{intro:recaled potential}
W_m:=m^{-1}\big(\nu-\rho_{Q_m}\big)\star|\cdot|^{-1}(x/m)
\end{equation}
converges weakly to the unique solution $W_{\nu}$ of Poisson's equation
\begin{equation}
-{\rm div}(\dem\nabla W_{\nu})=4\pi\,\nu.
\end{equation}
The matrix $\dem$ is the one which will appear in Pekar's theory below. 
The main reason for this is the fact, not derived in~\cite{CanLew-10} but proved in this paper, that the energy behaves as
\begin{equation}\label{eq:limit crys macro}
\lim_{m\to0}m^{-1} F_{\rm crys}\big[m^3\nu(m\cdot)\big]=2\pi \int_{\R^3}|\widehat{\nu}(k)|^2\left(\frac1{k^T\dem k}-\frac1{|k|^2}\right)dk,
\end{equation}
for any fixed $\nu$, see Theorem \ref{theo:main crys} below.

\subsection{Pekar's polarons in an anisotropic continuous medium}\label{sec:generalized-Pekar}
In this section we introduce the Pekar functionals which will describe the macroscopic behavior of our ($N$-)polaron. Since we want to consider anisotropic crystals, we have to generalize the formulas quoted in introduction~\eqref{eq:Pekar-intro} and~\eqref{eq:Pekar-N-intro} to the case of $\dem$ being a $3\times3$ symmetric matrix. In the macroscopic limit considered later, the matrix $\dem>1$ will be that of the reduced Hartree-Fock crystal derived in~\cite{CanLew-10} and recalled in the previous section.

We start with Pekar's generalized functional which we define as
\begin{equation}\label{ptgf1}
\ptgf [\psi] :=  \frac{1}{2}\int_{\R ^{3}} |\nabla \psi| ^2 dx  +  \ptgint \big[|\psi|^2\big]
\end{equation}
with
\begin{equation}\label{ptgint}
 \ptgint [\rho]: = 2\pi \int_{\R^3} \left| \hat{\rho} (k)\right| ^2 \left(\frac{1}{k^T \dem k} - \frac{1}{|k| ^2} \right)  \,dk.
\end{equation}
Alternatively we can define $\ptgf$ and 
$\ptgint$ by considering the solution $W_\rho$ of Poisson's equation
\begin{equation}\label{poteff}
-\nabla \left( \dem \nabla \Wrho \right) = 4 \pi \rho.
\end{equation}
We then have
\begin{equation}\label{ptgf1bis}
\ptgf [\psi] :=  \frac{1}{2}\int_{\R ^{3}} |\nabla \psi| ^2 dx  + \frac{1}{2} \int_{\R^3} |\psi|^2 \left( W_{|\psi|^2} - |\psi|^2 \star |\cdot| ^{-1}\right).
\end{equation}
We define the corresponding ground state energy as
\begin{equation}\label{ptge}
\ptge(1) = \inf \left\lbrace \ptgf [\psi],\ \int_{\R ^3} |\psi| ^2 = 1  \right\rbrace.
\end{equation}
The following deals with the existence of minimizers for the variational problem~\eqref{ptge}.

\begin{theorem}[\textbf{Ground states of Pekar's functional, $N=1$, \cite{Lieb-77,Lions-84}}]\label{theo:PTg1}\mbox{}\\
We assume that $\dem>1$ in the sense of symmetric matrices.
\begin{enumerate}
\item (\textbf{Existence}). The minimization problem~\eqref{ptge} admits at least one minimizer $\ptgm$. It satisfies the nonlinear equation
\begin{equation}\label{equationptg1}
-\frac{1}{2} \Delta \ptgm + \left( W_{|\ptgm|^2}-\left| \ptgm \right| ^2 \star | \cdot | ^{-1} \right) \ptgm
= \lambda\; \ptgm.
\end{equation}
\item (\textbf{Convergence of minimizing sequences}). All the minimizing sequences $(\psi_n)$ for $\ptge (1)$ are precompact in $H ^1 (\R ^3)$, 
up to a translation. That is, there exists a sequence $(\tau_k)\subset\R^3$ and a minimizer $\ptgm$ of $\ptge(1)$ such that $\psi_{n_k}(\cdot-\tau_k)\to\ptgm$ strongly in $H^1(\R^3)$.
\end{enumerate}
\end{theorem}

This result is a standard statement on which we will not elaborate. It can be proved by using well known techniques of nonlinear analysis, which are similar to what we do later in Section \ref{sec:macro 3}. When $\dem$ is proportional to the identity, Theorem~\ref{theo:PTg1} was proved first by Lieb in~\cite{Lieb-77}, and in this special case the minimizer is unique (up to translations). Other functionals very similar to $\ptgf$ have been considered by Lions in~\cite{Lions-84} for $N=1$.
To our knowledge, uniqueness for an anisotropic $\dem$ is not known.

Of course, the main reason why $\ptge(1)$ always has a minimizer is that the resulting potential is attractive at long distances. This is the case even when $\dem$ has 1 as an eigenvalue, as soon as $\dem\neq1$ (we have $k^T\dem k>|k|^2$ except on a set of measure zero).

\medskip

We now turn to Pekar's multi-polaron problem, whose energy functional in an anisotropic medium is
\begin{multline}\label{ptgfN}
\cE^{\rm P}_{\varepsilon_{\rm M}}[\Psi]=\int_{\R^{3N}}\left(\frac12\sum_{j=1}^N|\nabla_{x_j}\Psi(x_1,...,x_N)|^2+\sum_{1\leq k<\ell\leq N}\frac{|\Psi(x_1,...,x_N)|^2}{|x_k-x_\ell|}\right)dx_1\cdots dx_N\\  + \ptgint [\rhoP]
\end{multline}
where $\ptgint$ is defined above in~\eqref{ptgint} and with $\rho_\Psi$ the density of the many-body wave function $\Psi$ (see \eqref{defi densite pre}).
The corresponding ground state energy is obtained by minimizing this functional amongst antisymmetric $N$-body wave functions :
\begin{equation}\label{ptgeN}
\ptge (N)= \inf \left\lbrace \ptgf[\Psi],\: \int_{\R ^{3N}} |\Psi| ^2 = 1, \: \Psi \mbox{ antisymmetric }  \right\rbrace.
\end{equation}
We recall that an $N$-body wave function is antisymmetric if
\begin{equation}\label{eq:fermionic}
\Psi (x_1,\ldots,x_i,\ldots,x_j,\ldots, x_N) = -\Psi (x_1,\ldots,x_j,\ldots,x_i,\ldots, x_N) \quad \forall i\neq j.
\end{equation}

For the $N$-polaron there is a competition between the electronic repulsion and the effective polaronic interaction $F_{\rm crys}$.
There does not always exist minimizers to the variational problem~\eqref{ptgeN}. Indeed, Frank, Lieb, Seiringer and Thomas have proved in~\cite{FraLieSeiTho-10b,FraLieSeiTho-10a} that when $\varepsilon_{\rm M}\leq 1+a$ (with $a$ independent of $N$), $\cE^{\rm P}_{\varepsilon_{\rm M}}$ has no bound state when $N\geq2$. The following theorem deals with the question of existence. 

\begin{theorem}[\textbf{Ground states of Pekar's functional, $N\geq2$,~\cite{Lewin-11}}]\label{theo:PTgN}\mbox{}\\
 We assume that $\dem>1$ in the sense of symmetric matrices.
The following assertions are equivalent
\begin{enumerate}
\item (\textbf{Binding}). One has 
\begin{equation}\label{binding ptgN}
\ptge (N) < \ptge (N-k) + \ptge (k) \mbox{ for all } k=1 \ldots N-1. 
\end{equation}

\smallskip

\item (\textbf{Convergence of minimizing sequences}). All the minimizing sequences for $\ptge (N)$ are precompact in $H ^1 (\R ^{3N})$, up to a translation. In particular, there exists a minimizer $\Psi^{\rm P}_{\dem}$ for $\ptge(N)$, solving the many-body nonlinear Schrödinger equation
\begin{equation}
\!\!\!\!\left(\sum_{j=1}^N\left(-\frac12\Delta_{x_j} +W_{\rho_{\Psi^{\rm P}_{\dem}}}(x_j)-\rho_{\Psi^{\rm P}_{\dem}}\star|\cdot|^{-1}(x_j)\right)+\sum_{1\leq k<\ell\leq N}\frac{1}{|x_k-x_\ell|}\right)\Psi^{\rm P}_{\dem}=\lambda\; \Psi^{\rm P}_{\dem}.
\end{equation}
\end{enumerate}
Furthermore, for every fixed $N$ there exists a constant $c_N<\ii$ such that~\eqref{binding ptgN} is satisfied for all $\dem> c_N$.
\end{theorem}

The above theorem is a particular case of Theorem 25 in \cite{Lewin-11} (see also Theorem 28 and Remark 15 therein).
Frank, Lieb and Seiringer have shown recently~\cite{FraLieSei-11} that there is also a minimizer if $\dem$ is the limit of dielectric matrices $\varepsilon_n\to\dem$ for which~\eqref{binding ptgN} is valid for all $n$.
The binding inequality~\eqref{binding ptgN} is a necessary condition for the compactness of minimizing sequences, but not for the existence of minimizers.

%%%%%%%%%%%%%%%%%%%%%%%%%%%%%%%%%%%%%%%%%%%%%%%%%%%%%%
\subsection{Derivation of Pekar's $N$-polaron}
In this section we state our main results on the convergence of the polaron in the macroscopic limit. 
As motivated in the introduction, we define the energy functional for an electron interacting with a microscopic quantum crystal by
\begin{equation}
\cE_m[\psi]:=\frac1{2}\int_{\R^3}|\nabla\psi(x)|^2\,dx+m^{-1}\int_{\R^3}V^0_{\rm per}(x/m)|\psi(x)|^2\,dx+m^{-1}F_{\rm crys}\big[m^{3}|\psi(m\cdot)|^2\big],
\label{totf1}
\end{equation}
with $V^0_{\rm per}$ and $F_{\rm crys}$ as in Section~\ref{sec:model-crystal}. The corresponding ground state energy is 
\begin{equation}
E_m(1)=\inf \left\lbrace \cE_m [\psi],\ \int_{\R ^3} |\psi| ^2 = 1  \right\rbrace.
\label{tote1}
\end{equation}
Similarly, for $N\geq2$ the $N$-polaron energy is defined as
\begin{multline}
\cE_{m}[\Psi]=\int_{\R^{3N}}\left(\frac12\sum_{j=1}^N|\nabla_{x_j}\Psi(x_1,...,x_N)|^2+\sum_{1\leq k<\ell\leq N}\frac{|\Psi(x_1,...,x_N)|^2}{|x_k-x_\ell|}\right)dx_1\cdots dx_N\\
+m^{-1}\int_{\R^3}V^0_{\rm per}(x/m)\,\rho_\Psi(x)\,dx+m^{-1}F_{\rm crys}\big[m^{3}|\rhoP(m\cdot)|^2\big],
\label{totfN}
\end{multline}
and the associated ground state energy is
\begin{equation}\label{toteN}
\tote(N) := \inf \left\lbrace \cE_m [\Psi], \: \int_{\R ^{3N}} |\Psi| ^2 = 1,\: \Psi \mbox{ antisymmetric}  \right\rbrace.
\end{equation}

Similarly to Theorems~\ref{theo:PTg1} and~\ref{theo:PTgN}, it can be shown~\cite{LewRou-11b} that there always exists at least one minimizer to the small polaron variational problem~\eqref{tote1}, and that binding inequalities imply the existence of a ground state for the $N$-polaron problem~\eqref{toteN}. Here we do not need this and we simply introduce the concept of approximate minimizers.

\begin{definition}[\textbf{Sequence of approximate minimizers}]\label{def:minimiseur}\mbox{}\\
\it
Let $N\geq1$. We say that a sequence $\left( \Psi_m \right)$ of $L^2(\R^{3N})$-normalized antisymmetric $N$-body wave functions is a sequence of \emph{approximate minimizers} for $\tote(N)$ if
\begin{equation}\label{minimiseur 1}
\lim_{m\to0}\left(\cE_m[\Psi_m]-E_m(N)\right)=0.
\end{equation}
\end{definition}

In order to properly state our main result, we need to introduce an $\cL$-periodic function $u^{\rm per}_m$ which will describe the fast oscillations of our polaron, at the microscopic scale.

\begin{definition}[The function $u^{\rm per}_m$]\label{def:u_per}\mbox{}\\
\it
We denote by $u^{\rm per}_m$ the unique positive solution of  
\begin{equation*}
\Epere = \inf \left\lbrace\Eperf [v] \: :\: v\in H^1_{\rm per}(\Gamma), \int_{\Gamma} |v| ^2 = |\Gamma| \right\rbrace = \Eperf [\Eperm] 
\end{equation*}
with 
\begin{equation}\label{eq:Eperf defi}
\Eperf [v] = \int_{\Gamma} \frac{1}{2m} |\nabla v| ^2 + \FSp |v| ^2.
\end{equation}
By periodicity $\Eperm$ is extended to the whole of $\R^3$.
\end{definition}
In Section~\ref{sec:periodic} below, we will show using first-order perturbation theory that 
$u^{\rm per}_m\to1$ in $L^\ii(\R^3)$ when $m\to0$,  and that
\begin{equation}
\Eperelim:=\lim_{m\to0} m ^{-1} \Epere = \lim_{m\to0}m^{-1}\Eperf[\Eperm] = \int_{\Gamma} V^0_{\rm per} f^{\rm per}.
\label{eq:def_E_per}
\end{equation}
The function $f^{\rm per}\in L^\ii(\R^3)$ is the unique $\cL$-periodic solution to 
\[
\begin{cases}
\Delta f^{\rm per}=2V^0_{\rm per} \\
\int_{\Gamma}f^{\rm per}=0.
\end{cases}
\]
It appears in the perturbative expansion of $u^{\rm per}_m$:
\begin{equation}\label{eq:uper expansion}
\left\|u^{\rm per}_m-1-m f^{\rm per}\right\|_{L^{\infty} (\R ^3)}\leq Cm^2.
\end{equation}
The following is our main result on the behavior of approximate minimizers for $E_m(N)$ in the macroscopic limit $m\to0$.

\begin{theorem}[\textbf{Derivation of Pekar's $N$-polaron, $N\geq1$}]\label{theo:main}\mbox{}\\
We denote by $\dem>1$ the electronic dielectric matrix which was derived in~\cite{CanLew-10}.
Let $N\geq1$ be any positive integer. 
\begin{itemize}
\item\textbf{(Energy asymptotics)} We have
\begin{equation}\label{resulte1}
\boxed{\lim_{m\to0}E_m(N)=N\,\Eperelim + \ptge(N)}
\end{equation}
where we recall that $\Eperelim$ is the periodic ground state energy defined in~\eqref{eq:def_E_per}, and $\ptge(N)$ is Pekar's ground state energy defined in~\eqref{ptge} for $N=1$ and in~\eqref{ptgeN} for $N\geq2$.

\bigskip

\item \textbf{(Convergence of states)} Let $(\Psi_m)_m$ be a sequence of approximate minimizers for $E_m(N)$, in the sense of Definition \ref{def:minimiseur}. We define $\Psi^{\rm pol}_m$ by the relation
\begin{equation}\label{minimiseur pol}
\boxed{\Psi_m (x_1,...,x_N)= \prod_{j=1}^N\Eperm (x_j/m)\;\, \Psi^{\rm pol}_m (x_1,...,x_N).}
\end{equation}
Then $(\Psi^{\rm pol}_m)_m$ is a minimizing sequence for Pekar's variational problem $\ptge(N)$. 

If $N=1$ or if $N\geq2$ and the binding inequality~\eqref{binding ptgN} is satisfied, there exists a sequence of translations $(\tau_m)\subset\R^3$ and a minimizer $\Psi^{\rm P}_{\dem}$ of $\ptge(N)$ such that
\begin{equation}\label{state converge}
\Psi^{\rm pol}_m (x_1-\tau_m,...,x_N-\tau_m) \to \Psi^{\rm P}_{\dem}(x_1,...,x_N) \mbox{ strongly in } H^1 (\R ^3)^N  
\end{equation}
along a subsequence when $m\to 0$.
\end{itemize}
\end{theorem}

\begin{remark}
For this result, the fermionic nature of the electrons is not essential. The same theorem holds if the wave function $\Psi$ is supposed to be symmetric, i.e. if the electrons are replaced by bosons.
\end{remark}

From~\eqref{minimiseur pol}, we see that the polaronic wave function $\Psi_m$ has a multiscale behavior. The state is at the largest scale described by Pekar's polaron function $\Psi^{\rm P}_{\varepsilon_{\rm M}}$ which only depends on the macroscopic dielectric constant $\varepsilon_{\rm M}$ of the crystal. It also has a fast oscillatory behavior encoded in the $(m\cL)$-periodic term $u_m^{\rm per}(x/m)$. 
According to \eqref{eq:uper expansion}, this factor tends to $1$ in $L^\ii(\R^3)$ when $m\to0$, but it contributes to $\nabla\Psi_m$ and to the kinetic energy, yielding the term $N\,E^{\rm per}$ in the total energy. The function $u_m^{\rm per}$ depends on the detailed microscopic structure of the crystal via the electrostatic periodic potential $V^0_{\rm per}$ or, equivalently, the nuclear and electronic densities $\mu^0_{\rm per}$ and $\rho^0_{\rm per}$.

\medskip

In the theory of homogenization, one often multiplies the microscopic periodic potential by $m^{-2}$ instead of our $m^{-1}$ (see, e.g.,~\cite{AllPia-05,Sparber-06}). The physical motivations leading to the factor $m^{-1}$ in front of $V^0_{\rm per}$ automatically place us in a perturbative regime for the periodic equation~\eqref{eq:equation_u_per}. Actually, we could replace $u^{\rm per}_m(x/m)$ in~\eqref{minimiseur pol} by its first-order approximation $1+m\,f^{\rm per}(x/m)$ without changing the result.

For this reason, we need not use elaborate tools to analyze the contribution of the fast oscillations to the ground state. A simple energy decoupling method allows to separate the energy into the contribution of the oscillations and a functional where the fast potential $\FSp$ is absent but $\Eperm$ appears as a weight. The main point is that \emph{no derivatives} of $\Eperm$ appear in the latter functional. Then \eqref{eq:uper expansion} ensures that one can ignore the weight $\Eperm$ and simply consider a small polaron functional where the potential $\FSp$ has disappeared, i.e. the two scales of the problem completely decouple. The details of this procedure are provided in Section \ref{sec:periodic}.

\medskip

The main difficulty of our work is to deal with the highly nonlinear functional $F_{\rm crys}$ which is at the origin of the dielectric matrix $\dem$ and of the possible binding of polarons. Our main contribution in this direction is the following result, in which we show that, in a macroscopic regime, $F_{\rm crys}$ can be replaced by $F^{\rm P}_{\dem}$ under fairly general assumptions.

\begin{theorem}[Macroscopic behavior of $F_{\rm crys}$]\label{theo:main crys}\mbox{}\\
Let $\boldsymbol{\psi}=(\psi_m)$ be a bounded sequence in $H^s(\R^3)$ for some $s>1/4$. Then, we have
\begin{equation}\label{eq:limit crys}
\lim_{m\to0}\Big(m^{-1}F_{\rm crys}\big[m^3|\psi_m(m\cdot)|^2\big]-F^{\rm P}_{\dem}\big[|\psi_m|^2\big]\Big)=0.
\end{equation}
\end{theorem}

\medskip

It is rather easy to deduce our main Theorem~\ref{theo:main} from this result (with $s=1$), once the fast oscillations have been removed from the energy. The argument is provided in Section~\ref{sec:completion} below.

Theorem \ref{theo:main crys} is based on a perturbative expansion of the energy in the macroscopic limit. In the case where $\psi_m\equiv\psi$ is a fixed function, perturbation theory was used in \cite{CanLew-10} to compute the reaction of the Fermi sea of the reduced Hartree-Fock crystal on first order (this is the content of Theorem 3.1 of \cite{CanLew-10}, recalled above). The corresponding leading order of the energy was not computed however, the missing ingredient being an asymptotic expression of the kinetic energy that we provide in Section~\ref{sec:aux_interaction} below.  As we will see, \eqref{eq:limit crys} easily follows from this calculation when $\psi_m\equiv\psi$. 

When $\psi_m$ depends on $m$, then \eqref{eq:limit crys} is much more subtle. The Pekar interaction $\ptgint$ is an electrostatic interaction (similar to a simple Coulomb term) and, by the Hardy-Littlewood-Sobolev inequality~\cite{LieLos-01}, it can be controlled by the $L^{12/5}(\R^3)$ norm of $\psi_m$. By~\eqref{eq:lower-bound-F-crys}, the same is true for the crystal interaction $\crysint$. It is therefore natural to assume that $(\psi_m)$ is bounded in $L^{12/5}(\R^3)$. However, for the macroscopic limit \eqref{eq:limit crys} to be true, it is also crucial that $\psi_m$ stays spread over a macroscopic region. The sequence $(\psi_m)$ could be bounded in $L^{12/5}(\R^3)$ and still concentrate on a scale of order $m$. A counter example to \eqref{eq:limit crys} when 
$(\psi_m)$ is only bounded in $L^{12/5}(\R^3)$ is provided later in Section \ref{sec:macro 2}, Eq.~\eqref{eq:counter-example}.

The role of our assumption that $(\psi_m)$ is bounded in $H^s(\R^3)$, is precisely to ensure that $(\psi_m)$ stays \emph{locally compact} in $L^{12/5}(\R^3)$ and does not blow up at a scale of order $m$, by the Sobolev inequality and the Rellich-Kondrachov theorem. The constraint $s>1/4$ is necessary to have that $12/5$ is subcritical, that is, below the Sobolev exponent $6/(3-2s)$.\\ 

The proof of Theorem \ref{theo:main crys} being the technical core of our paper, it is worth outlining its main steps : 
\begin{itemize}
\item In Section \ref{sec:aux_interaction}, we show by perturbation theory that one can replace the exact density matrix $\gamma_m=\gamma^0_{\rm per}+Q_m$ of the crystal by its first order approximation, without changing $\crysint$ to leading order. More precisely we write 
\[
Q_m = Q_1 + R_2
\] 
where $Q_1 = O(m)$ and $R_2 = O(m ^{3/2})$ in an appropriate norm, allowing to take into account only $Q_1$ when computing the leading order of the energy. This leads to an auxiliary functional $\auxint$ where the effect of the polarizable medium is expressed through a complicated but explicit operator $\mathcal{K}$ studied in~\cite{CanLew-10} and defined below. This step relies on the application of the Cauchy formula and a resolvent expansion to \eqref{eq:rHF-defect}, complemented with quite a bit of algebra using the Bloch-wave decomposition of $\FSh$.

\smallskip

\item  We then prove in Section \ref{sec:macro 2} that \eqref{eq:limit crys} holds, when $\psi_m\to\psi$ strongly in $L^{12/5}(\R^3)$. Using the perturbative result of the previous step, this is essentially an application of Theorem 3.1 in~\cite{CanLew-10}.

\smallskip

\item We finally consider the general case. In order to quantify the intuitive fact that a sequence $(\psi_m)$ which is bounded in $H^s(\R^3)$ has to stay spread over a macroscopic region, we use a so-called \emph{bubble decomposition}. This means that we split $\psi_m$ into a sum of pieces receding from each other in space and converging strongly in $L^{12/5}(\R^3)$, plus a rest that is small in $L^{12/5}(\R^3)$, using standard tools of nonlinear analysis as developed by Lieb~\cite{Lieb-83}, Lions~\cite{Lions-84,Lions-84b} and others. The goal is now to show that both terms $m^{-1}F_{\rm crys}\big[m^3|\psi_m(m\cdot)|^2\big]$ and $F^{\rm P}_{\dem}\big[|\psi_m|^2\big]$ are equal to the sum of the energies of each piece, up to a controlled error. Since each piece of mass converges strongly in $L^{12/5}(\R^3)$, the previous step then implies that the difference between these two sums is small when $m\to0$. This technical argument is detailed in Sections \ref{sec:macro 2} and \ref{sec:macro 3}.
\end{itemize}

Clearly, Theorem \eqref{theo:main crys} cannot follow from simple weak limit arguments because of the translation-invariance properties of $\crysint$ and $\ptgint$. The term involving $\crysint$ does not change if $\psi_m$ is replaced by $\psi_m(\cdot-\tau_m)$ with $\tau_m$ in the rescaled lattice $m\cL$, whereas $\ptgint$ is itself fully translation-invariant. In particular, one can easily construct a sequence of functions such that $\psi_m \rightharpoonup \psi$ (e.g. in $H^1 (\R ^3)$), but with  
\[
\lim_{m\to0} m^{-1}F_{\rm crys}\big[m^3|\psi_m(m\cdot)|^2\big] \neq F^{\rm P}_{\dem}\big[|\psi|^2\big].
\]
The natural counter-example is of the form $\psi_m = \psi + \varphi (.-\tau_m)$ where $\psi$ and $\varphi$ are fixed compactly supported functions and $\tau_m$ is a translation, $|\tau_m| \to \infty$ when $m\to 0$. It is thus clear that our method of proof must in particular accommodate such type of behaviors and this is precisely what the bubble decomposition does.

\begin{remark}
The sign of the macroscopic density $\nu_m=|\psi_m|^2\geq0$ considered in Theorem~\ref{theo:main crys} was only motivated by our application (the polaron). As can be seen from our method of proof, the same result remains true if $(|\psi_m|^2)$ is replaced by a (real-valued) sequence $(\nu_m)$ with no particular sign, but with appropriate Sobolev bounds ensuring strong local compactness in $L^{6/5}(\R^3)$.
\end{remark}

The rest of the paper is devoted to the proof of our main results, Section \ref{sec:periodic} dealing with the periodic problem and the associated energy decoupling, Section \ref{sec:effective-interaction} with the proof of Theorem \ref{theo:main crys}. The final steps in the proof of Theorem \ref{theo:main} are presented in Section \ref{sec:completion}.

\bigskip

\noindent \textbf{Notation : } In the whole paper, $C$ denotes a generic positive constant whose value may change from line to line.

%%%%%%%%%%%%%%%%%%%%%%%%%%%%%%%%%%%%%%%%%%%%%%%%%%%%%%
%%%%%%%%%%%%%%%%%%%%%%%%%%%%%%%%%%%%%%%%%%%%%%%%%%%%%%
\section{The microscopic oscillations and scale decoupling}\label{sec:periodic}
As we have explained, our polaron has a multiscale structure. The macroscopic scale is described by Pekar's theory, which only depends on the macroscopic dielectric tensor $\varepsilon_{\rm M}$. The microscopic oscillations are described by a simple linear eigenvalue problem which we study in this section. We also explain here how to decouple these two scales.

\bigskip

We introduce an energy functional where $\Eperm$ (that we extend to the whole of $\R^3$ by periodicity) appears as a weight. In the case $N=1$, it is
\begin{equation}\label{startf1h}
\htotf [\psi] = \frac{1}{2} \int_{\R ^3}\left| \Eperm (x/m) \right| ^2 |\nabla \psi(x)| ^2\,dx + m ^{-1}\crysint \left[ m ^3 |\Eperm| ^2 |\psi (m\cdot)| ^2 \right].
\end{equation}
In the case $N\geq 2$ we introduce
\begin{equation}\label{EperM}
\EperM (x_1 \ldots, x_N) := \prod_{i=1} ^N \Eperm (x_i)
\end{equation}
and the corresponding weighted functional is now
\begin{equation}\label{startfNh}
\htotf [\Psi] = \sum_{j=1} ^N \frac{1}{2} \int_{\R ^{3N}}\left| \EperM (\cdot/m) \right| ^2 |\nabla_j \Psi| ^2 +\sum_{1\leq k<\ell\leq N}\frac{|\EperM(\cdot/m) \Psi|^2}{|x_k-x_\ell|}+ m ^{-1}\crysint \left[ m ^3 \trhoP (m\cdot) \right]
\end{equation}
with the modified density
\begin{equation}\label{eq:rhoP tilde}
\trhoP (x):= \left|\Eperm (x/m)\right| ^2 \int_{\R^{3(N-1)}} \prod_{i=2} ^N \left| \Eperm(x_i) \right| ^2 \left|\Psi(x,x_2,\ldots,x_N)\right| ^2 dx_2 \ldots dx_N.
\end{equation}

The link between these energy functionals and our original problem will be made clearer in Lemma \ref{lem:decouple}.

As mentioned above, the two scales appearing in our problem, namely that of the crystal lattice and that of the minimizer of the generalized Pekar functional decouple. Mathematically this means that the factors involving $\Eperm$ in the ``tilde'' functional introduced above can be ignored because $\Eperm \approx 1$ on the macroscopic scale. This is demonstrated in the 
\begin{lemma}[\bf The periodic problem]\label{lem:uper}\mbox{}\\
Let $\fper$ be the solution of 
\begin{equation}\label{defi fper}
\begin{cases}
\displaystyle
-\frac{1}{2} \Delta \fper = - \FSp \mbox{ in } \Gamma \\
\displaystyle \int_{\Gamma} \fper = 0 
\end{cases}
\end{equation}
with periodic boundary conditions on $\dd \Gamma $. There holds
\begin{equation}\label{estim uper}
\left\Vert \Eperm  - 1 - m\fper \right\Vert_{L ^{\infty} (\Gamma)} \leq C m ^2 
\end{equation}
and
\begin{equation}\label{estim Eper}
\Epere = m \int_{\Gamma} \FSp \fper + O (m^2). 
\end{equation}
\end{lemma}

\begin{proof}
The arguments here are standard so the the following is voluntarily kept at the level of a sketch. We first remark that $\FSp \in L ^{p} (\Gamma)$ at least for any $1 \leq p<3$ since it is defined as 
$-\Delta \FSp = 4 \pi \left( \FSn - \FSd \right)$
with $\FSn$ a non-negative finite measure, and $\FSd \in L ^2 (\Gamma)$.
On the other hand, simple upper and lower bounds yield $\Epere = O(m)$. From the equation
\[
-\frac{1}{2} \Delta\Eperm + m \FSp \Eperm = m \Epere \Eperm
\]
we then deduce that $\left\Vert \Eperm -1 \right\Vert_{H^1 (\Gamma)} = O(m)$. Writing 
$\Eperm = 1 + m f$
it is then easy to see that necessarily
$\int_{\Gamma} f = O(m)$
and
\[
-\frac{1}{2} \Delta f = \Epere - \FSp + O_{L^p}(m)
\]
for any $1\leq p< 2 $. This implies that $\left\Vert f - \fper\right \Vert _{L ^{\infty} (\Gamma)} \leq C m $ (recall that $\Epere = O(m)$) by standard elliptic regularity and the conclusion of the Lemma follows.
\end{proof}

The following lemma is the key tool allowing us to effectively decouple the two different scales of the problem. Such decoupling techniques 
have already been used in several contexts, e.g. Ginzburg-Landau theory \cite{LasMir-99} and the derivation of Gross-Pitaevskii theory from many-body quantum physics \cite{LieSeiYng-00}.\\
\begin{lemma}[\textbf{Energy Decoupling}]\label{lem:decouple}\mbox{}\\
Let $\Psi \in H^1 (\R ^{3N})$ be a normalized wave function, $\int_{\R ^{3N}} |\Psi| ^2 = 1$.
Define $\FP$ by the formula
\begin{equation}\label{decouple func}
\Psi =  \EperM (\cdot/m)\,  \FP
\end{equation}
where $\Eperm$ is extended to $\R ^3$ by periodicity and $\EperM(x_1,...,x_N)=\prod_{j=1}^N\EperM(x_j)$. We have
\begin{equation}\label{decouple ener1}
\totf [\Psi] = N m^{-1} \Epere + \htotf [\FP].
\end{equation}
\end{lemma}

\begin{proof}
Let us first remark that $\FP$ is well defined by formula \eqref{decouple func}, because $\Eperm$ is strictly positive (it is the ground state of a Schr\"odinger operator).

We only write the proof for $N=1$. Using that $\trhoP = \rho_{\EperM (\cdot/m) \Psi}$, the argument is exactly the same for $N\geq2$.
The only input of the lemma is the variational equation satisfied by $\Eperm$: 
\begin{equation}\label{Eperm equation}
-\frac{1}{2}\Delta \Eperm + m \FSp \Eperm = m \Epere \Eperm.
\end{equation}
From the definition \eqref{decouple func} and an integration by parts, we have
\[
\int_{\R ^3} \left| \nabla \Psi \right| ^2 =  m ^{-2} \int_{\R ^3} \left( - \Delta \Eperm (\cdot/m) \right) \Eperm (\cdot/m)|\FP| ^2 + \int_{\R ^3} |\Eperm (\cdot/m)| ^2 |\nabla \FP| ^2.
\]
Hence, using \eqref{Eperm equation} and the normalization of $\Psi$, we obtain
\begin{multline*}
\frac{1}{2}\int_{\R ^3} \left| \nabla \Psi \right| ^2 + m ^{-1} \int_{\R ^3} \FSp (\cdot/m) |\Psi| ^2 + \crysint\big[m ^3|\Psi (m\cdot)| ^2\big]
\\ = m ^{-1} \Epere + \int_{\R ^3} |\Eperm (\cdot/m)| ^2 |\nabla \FP| ^2 + \crysint\big[m ^3 | \Eperm | ^2 |\FP (m\cdot)| ^2\big].
\end{multline*}
This gives the result in the case $N=1$.
\end{proof}

\section{Derivation of Pekar's interaction: proof of Theorem~\ref{theo:main crys}}\label{sec:effective-interaction}
In this section we provide the proof of Theorem~\ref{theo:main crys}, which is our main result relating, in the macroscopic limit,
the exact crystal interaction $\crysint$ to the generalized Pekar interaction $\ptgint$.

We start by recalling in Section \ref{sec:input} several properties of the crystal model that we use in our analysis. In particular we introduce an auxiliary model that is `halfway' between the full interaction via the crystal and the Pekar macroscopic interaction. In Section \ref{sec:aux_interaction} we explain how to go from the crystal model to the auxiliary model. The latter is a Coulomb-type interaction involving a complicated but explicit operator denoted $\K$. This operator $\K$ is nothing else but the linear response of the crystal in the sense that, if $\rho_Q$ denotes the perturbation of the crystal density in presence of the defect $\nu$, one has $\rho_Q\simeq -\K\nu$ for $\nu$ small enough (see~\eqref{eq:1st-order} below). Results about $\K$ proved in \cite{CanLew-10} (alluded to briefly in the introduction and recalled below) suggest that the intermediary model naturally turns into a Pekar theory in the limit. Namely, they allow, in combination with the analysis 
 of Section \ref{sec:aux_interaction}, to prove \eqref{eq:limit crys macro}. However, the proof of Theorem \ref{theo:main crys} requires to implement these ideas under much more general assumptions than was done in \cite{CanLew-10}. This is done by using a bubble decomposition for the sequence $(\psi_m)$, which is the content of Sections \ref{sec:macro 2} and \ref{sec:macro 3}.

\subsection{Useful properties of the crystal}\label{sec:input}

In order to realize our program, we first recall several useful facts about the polarized crystal in an external density $\nu$, whose proofs can be found in \cite{CanDelLew-08a,CanLew-10}. 

\subsubsection{Linear response of the Fermi sea}

We denote by
\begin{equation}\label{eq:crystotf}
\cryse [Q,\nu] = D(\nu,\rho_Q)+\cF_{\rm crys}[Q]
\end{equation}
the total energy of the crystal in presence of the defect $\nu$, with $\cF_{\rm crys}[Q]$ as in~\eqref{eq:def_energy_crystal}, and we recall that 
\begin{equation}
\crysint[\nu]=\inf_{- \FSm \leq Q \leq 1 -\FSm}\cryse[Q,\nu]. 
\label{eq:crystotf_min}
\end{equation}
The proper functional setting for the minimization problem \eqref{eq:crystotf_min} is detailed in \cite{CanDelLew-08a}, we only sketch it here. Any operator $Q$ satisfying the constraint 
\begin{equation}\label{eq:contrainte cristal}
- \FSm \leq Q \leq 1 -\FSm
\end{equation}
is decomposed as
\begin{equation}\label{eq:decomp Q}
Q = \Qmm + \Qmp +\Qpp + \Qpm 
\end{equation}
where $\Qmm = \FSm Q \FSm$, $\Qmp = \FSm Q \left(1-\FSm \right)$, and so on. It is proved in \cite{CanDelLew-08a} that for $Q$ satisfying \eqref{eq:contrainte cristal} and $\nu\in L^1(\R^3)\cap L^2(\R^3)$, $\cryse [Q,\nu]$ is finite if and only if $Q$ is in the function space 
\begin{equation}\label{eq:crys space}
\Q = \left\lbrace Q \in \Sch ^2  \Big| Q=Q^*,\: |\nabla |Q \in \Sch ^2, \: \Qpp, \Qmm \in \Sch ^1, \: |\nabla | \Qpp |\nabla |,\: |\nabla | \Qmm |\nabla |  \in \Sch ^1\right\rbrace.
\end{equation}
The space $\Q$ is endowed with its natural norm
\[
\Vert Q \Vert_{\Q} = \Vert Q \Vert_{\Sch ^2} + \Vert  \Qpp \Vert_{\Sch ^1 }+ \Vert \Qmm \Vert_{\Sch^1 } + \Vert  |\nabla| Q \Vert_{\Sch ^2 } +\Vert |\nabla| \Qpp |\nabla| \Vert_{\Sch ^1} + \Vert |\nabla| \Qmm |\nabla| \Vert_{\Sch ^1}.
\]
The symbols $\Sch^1$ and $\Sch  ^2$ denote the Schatten classes of trace-class and Hilbert-Schmidt operators on $L^2 (\R ^3)$ respectively (see~\cite{Simon-79} and~\cite{ReeSim4}).

An important fact is that, although the operators in $\Q$ are not necessarily trace-class, they are always locally trace-class and thus have an unambiguously defined associated density $\rho_Q\in L^1_{\rm loc}(\R^3)$. In Proposition~1 of \cite{CanDelLew-08a}, it is even shown that $\rhoQ$ belongs to $L^2(\R^3)$ and to the Coulomb space
\begin{equation}\label{eq:coul space}
\coul = \left\lbrace \rho \:\Big| \: D(\rho,\rho ) ^{1/2} <\infty \right\rbrace.
\end{equation}
By definition, there holds
\begin{equation}\label{eq:defi rhoQ}
\Tro (V Q ) = \int_{\R^3} V \rhoQ
\end{equation}
for any $V \in \coul'$ (the generalized trace $\Tro$ is defined in \eqref{eq:gen trace}). Moreover the inequality
\begin{equation}\label{eq:Q control coul}
\norm{\rho_Q}_{L^2(\R^3)}+D(\rhoQ,\rhoQ ) ^{1/2} \leq  C \left\Vert Q \right\Vert_{\Q}
\end{equation}
holds uniformly for $Q\in\Q$, showing that the linear map $Q\in\Q\mapsto\rhoQ\in L^2(\R^3)\cap \coul$ is continuous.

\medskip

The existence of $Q\in \Q$ minimizing \eqref{eq:crystotf_min} for fixed $\nu$ is proved in \cite{CanDelLew-08a}, Theorem 2. For ease of notation we will not emphasize the dependence of the minimizer of \eqref{eq:crystotf_min} on $\nu$. Minimizers are found to satisfy the self-consistent equation
\begin{equation}\label{equation Q-}
 Q = \oneep \left( \FSh + \left( \rhoQ + \nu \right)\star |\: . \:| ^{-1} \right) - \oneep \left( \FSh \right) + \delta
\end{equation}
where $\delta$ is a finite rank self-adjoint operator with 
\[
\mathrm{Ran} (\delta) \subset \mathrm{Ker} \left( \FSh + \left( \rhoQ + \nu \right)\star |\: . \:| ^{-1} \right).
\]
On the other hand, there holds
\[
 0 \geq \cryse[Q,\nu] \geq \frac{1}{2} D(\rhoQ,\rhoQ) + D(\nu, \rhoQ) \geq D(\rhoQ,\rhoQ) ^{1/2}\left( \frac{1}{2} D(\rhoQ,\rhoQ) ^{1/2} - D(\nu, \nu) ^{1/2}\right)
\]
where the upper bound is obtained by using a trial state $Q\equiv 0$ and the lower bound by dropping the positive kinetic energy term and 
using the Cauchy-Schwarz inequality. We deduce the uniform estimate
\begin{equation}\label{coulomb test}
D(\rho_Q, \rho_Q) \leq 4 D(\nu,\nu). 
\end{equation}
An easy consequence of the bound \eqref{coulomb test} is that the difference between $\FSh$ and $ \FSh + \left( \rhoQ + \nu \right)\star |\: . \:| ^{-1} $ is controlled by $D(\nu,\nu)$, in the resolvent sense. In particular, if $D(\nu,\nu)$ is small enough (depending on $\FSl$, see \cite{CanLew-10} Lemma 5 for the short proof) then $\tr_0(Q)=0$ and $\FSl\notin\sigma(\FSh + ( \rhoQ + \nu )\star |\: . \:| ^{-1})$. Then $\delta \equiv 0$ in \eqref{equation Q-} and $Q$ is unique: 
\begin{equation}\label{equation Q}
Q = \oneep \left( \FSh + \left( \rhoQ + \nu \right)\star |\: . \:| ^{-1} \right) - \oneep \left( \FSh \right).
\end{equation}

Since $\FSl$ is not in the spectrum of the mean-field operator when $D(\nu,\nu)$ is small enough, it is possible to express $Q$ using Cauchy's formula. Let $\CC$ be some curve in the complex plane that encloses the spectra of both $\FSh$ and $ \FSh + \left( \rhoQ + \nu \right)\star |\: . \:| ^{-1} $
in $(-\infty,\FSl)$. Using the resolvent formula one can then decompose $Q$ as 
\begin{equation}\label{Q test}
Q = Q_1 + R_2
\end{equation}
with
\begin{equation}
Q_1 := \ointC \frac{1}{z-\FSh}V \frac{1}{z-\FSh},\quad 
R_2 := \ointC \frac{1}{z-\FSh-V}\left(V  \frac{1}{z-\FSh }\right)^2
\label{Q1_R2}
\end{equation}
and
\begin{equation}
V := \left(\rhoQ + \nu \right)\star |\cdot| ^{-1}. 
\label{Vm}
\end{equation}
We again refer to \cite{CanLew-10}, in particular Section 6.3, for a more precise discussion of these facts.
When $D(\nu,\nu)$ is small, $V$ is also small in an appropriate sense because of \eqref{coulomb test}. One can then prove (see the proof of Lemma 3 in \cite{CanLew-10}) that 
\begin{eqnarray}
\left\Vert Q_1 \right\Vert_{\Q} &\leq& C \left\Vert V \right\Vert_{L ^2 + \coul '} \leq C D(\nu,\nu) ^{1/2}, \nonumber \\
\left\Vert R_2 \right\Vert_{\Q} &\leq& C \left\Vert V \right\Vert_{L ^2 + \coul '} ^2 \leq C D(\nu,\nu), \label{reste test}
\end{eqnarray}
for all $D(\nu,\nu)$ small enough.

\medskip

The main idea is now to just discard the second-order term $R_2$ and only keep $Q_1$ in our energy functional, leading to the auxiliary interaction $\auxint$. This is done as follows. First we consider the first order linear operator
\begin{equation}
\LL (\mu) := -\rho_{Q_1(\mu)},\quad\text{ where }\ 
Q_1(\mu) := \ointC \frac{1}{z-\FSh} \left(\mu \star |\: \cdot \:| ^{-1} \right) \frac{1}{z-\FSh}.
\label{def:cL}
\end{equation}
It was proved in~\cite{CanLew-10} (Proposition 2) that $\Ll$ is a bounded non-negative self-adjoint operator on the Hilbert space $\coul$. Explicit formulas can be found for the operator $\Ll$ in \cite{CanLew-10} but they are not needed here. 
Assuming that $D(\nu,\nu)$ is small enough, we can rewrite the self-consistent equation as follows
$$\rho_Q=-\Ll(\rho_Q+\nu)+\rho_{R_2}$$
or, equivalently,
$$\rho_Q=-\left(1-(1+\Ll)^{-1}\right)\nu+(1+\Ll)^{-1}\rho_{R_2}.$$
Introducing the operator
\begin{equation}
\boxed{\K:=1-(1+\LL)^{-1},}
\label{defiF} 
\end{equation}
which is bounded and non-negative on $\coul$, we deduce from \eqref{eq:Q control coul} and \eqref{reste test} that
\begin{equation}
\norm{\rho_Q+\K\,\nu}_\coul\leq \norm{\rho_{R_2}}_\coul\leq C \norm{R_2}_\Q \leq C\,D(\nu,\nu)
\label{eq:1st-order}
\end{equation}
for $D(\nu,\nu)$ small enough. The leading order for $\rho_Q$ is therefore $-\K\,\nu$, which is interpreted as the linear response of the nonlinear Fermi sea.

We now define the auxiliary interaction $\auxint$ by retaining only second order terms in the energy, which corresponds to keeping only the first order terms for $Q$ (the first order energy automatically vanishes). The correct formula turns out to be
\begin{equation}
\boxed{\phantom{\int}\auxint[\nu]=-\frac12 D(\nu,\K\,\nu),\phantom{\int}}
\label{Ff}
\end{equation}
as we will prove in Section \ref{sec:aux_interaction} below.

\subsubsection{Macroscopic dielectric matrix}

Other results of \cite{CanLew-10} will prove useful in our context. We define the dilation operator $\mathcal{U}_m$ and its adjoint $\mathcal{U}_m^*$ in $L^2(\R^3)$ by
\begin{equation}\label{dilation}
( \mathcal{U}_m \nu ) (x) = m^3\,\nu(m x),\qquad  ( \mathcal{U}_m^* \nu ) (x) = \nu(x/m).
\end{equation}
We then introduce the potential
\begin{equation}\label{Bm}
B_m (\nu) = m ^{-1} \, \mathcal{U}_m^* \left( |\: . \:| ^{-1} \star \left(1 +\Ll \right)^{-1} (\mathcal{U}_m \, \nu)\right).
\end{equation}
The above formula means that given a charge-defect $\nu$, we first scale it using $\mathcal{U}_m$, apply the operator $\left(1 +\Ll \right)^{-1}$, consider the corresponding electrostatic potential (convolution with the Coulomb kernel) and then scale back the total result using $\mathcal{U}_m ^*$. An easy computation shows that
\begin{equation}
\boxed{m ^{-1} \auxint [m ^3 \nu(m\cdot)]= \frac{1}{2} \left( \int_{\R^3} B_m (\nu) \nu - D(\nu,\nu) \right).}
\label{eq:formula_F_aux}
\end{equation}

Using that $\Ll\geq0$ as an operator in the Hilbert space $\coul$ (Proposition 2 in \cite{CanLew-10}), implying that $(1+\Ll)^{-1}\leq1$, we obtain 
\begin{equation}
0\leq \int B_m(\nu)\,\nu=m^{-1}D(\mathcal{U}_m\nu,(1+\Ll)^{-1}\mathcal{U}_m\nu)\leq m^{-1}D(\mathcal{U}_m\nu,\mathcal{U}_m\nu)=D(\nu,\nu).
\label{eq:bound_B_m} 
\end{equation}
This proves that $B_m$ is continuous from $\coul$ to $\coul '$, with a norm independent of $m$:
\begin{equation}
\norm{B_m}_{\coul\to\coul'}\leq 1.
\label{eq:estim_norm_Bm}
\end{equation}
Moreover, from the analysis in Section 6.10 of \cite{CanLew-10}, we have for any fixed $\nu\in\coul$ that
\begin{equation}\label{limit Bm}
B_m (\nu) \rightharpoonup W_{\nu}
\end{equation}
weakly in $\coul '$, where $W_{\nu}$ is the dielectric potential depending on the dielectric matrix $\dem$, which we have defined above in \eqref{poteff} by
\begin{equation}\label{poteff_bis}
-\nabla \left( \dem \nabla W_\nu \right) = 4 \pi \nu.
\end{equation}
This result is a step of the proof of Theorem 3.1 in \cite{CanLew-10}, whose main conclusion was recalled in Section \ref{sec:model-crystal}. 

\medskip

The connection between the full interaction with the crystal and Pekar's model goes roughly speaking as follows :
\begin{eqnarray*}
 m ^{-1} \cryse [m ^3 \nu(m.)] &\approx& m ^{-1} \auxint [m ^3 \nu(m.)] \\
&=& \frac{1}{2} \left( \int_{\R^3} B_m (\nu) \nu - D(\nu,\nu) \right) \\
&\approx& \frac{1}{2} \left( \int_{\R^3} W_{\nu} \nu - D(\nu,\nu) \right) = \ptgint [\nu].
\end{eqnarray*}
Section \ref{sec:aux_interaction} is devoted to the proof the first `equality'. The second equality is a simple scaling. The convergence \eqref{limit Bm} then proves that the third `equality' is justified when $\nu$ does not depend on $m$. To allow the $m$ dependence specified in the assumptions of Theorem \ref{theo:main crys} we need other ingredients, that are provided in Section \ref{sec:macro 2} and \ref{sec:macro 3}.

\subsection{Perturbation theory and the auxiliary interaction $\auxint$}\label{sec:aux_interaction}

In this section we use perturbation theory to approximate the nonlinear reaction of the crystal (described by the operator $Q$), by its first order approximation, obtained by considering that the defect $\nu$ is small. This will allow us to replace the complicated nonlinear interaction energy $\crysint$ by its leading order $\auxint$.
For completeness, we consider an abstract defect $\nu$ (for instance in $L^{6/5}(\R^3)$) and we provide uniform error bounds. 

We will come back later to the case of a sequence $\nu_m=m^3|\psi_m(m\cdot)|^2$ with $(\psi_m)$ bounded in $H^s(\R^3)$, as is considered in the statement of Theorem~\ref{theo:main crys}. As was already explained and used in \cite{CanLew-10}, the correct norm measuring the size of a given defect $\nu$ is that given by the Coulomb term $D(\nu,\nu)$. Saying differently, one can prove that the terms in the perturbation series of $\rho_Q$ can all be controlled by powers of $D(\nu,\nu)$. Fortunately, we have 
\begin{equation}
D\big(m^3|\psi_m(m\cdot)|^2, m^3|\psi_m(m\cdot)|^2\big)=m\,D\big(|\psi_m|^2,|\psi_m|^2\big)\leq C\,m\norm{\psi_m}_{L^{12/5}(\R^3)}^4\leq C\, m,
\label{eq;bound_12_5_psi_m} 
\end{equation}
when $(\psi_m)$ is bounded in $H^s(\R^3)$ with $s>1/4$. The macroscopic limit $m\to0$ thus automatically places us in this perturbative regime, which is why we will be able to only keep the leading order term.

\medskip

The following proposition justifies the introduction of the auxiliary interaction functional by quantifying its difference with the full interaction with the crystal.

\begin{proposition}[\textbf{From $\crysint$ to $\auxint$ via perturbation theory}]\label{pro:to intermediary}\mbox{}\\
There exist two positive constants $C$ and $\eta$ such that for any $\nu\in \coul$ with $D(\nu,\nu) \leq \eta$, we have 
\begin{equation}
\boxed{\phantom{\int}\Big| \crysint[\nu] - \auxint[\nu] \Big|\leq  C\, D(\nu,\nu) ^{3/2}\phantom{\int}} \label{upbound decouple}
\end{equation}
where $\auxint$ was defined above in~\eqref{Ff}.
\end{proposition}

\medskip

The rest of this section is devoted to the proof of Proposition \ref{pro:to intermediary}.
In the first step we rewrite the kinetic energy following~\cite{BacBarHelSie-99,HaiLewSer-05a}.

\begin{lemma}[\textbf{Rewriting the kinetic energy term}]\label{lem: calcul sup 1}\mbox{}\\
Under the assumptions of Proposition \ref{pro:to intermediary}, we have
\begin{equation}\label{formule Q}
\Qpp - \Qmm = Q ^2,
\end{equation}
and consequently
\begin{equation}\label{astuce energie 1}
\Tro \left( \left(\FSh - \FSl \right) Q \right) = \Tr \left( \left| \FSh-\FSl \right| Q ^2\right).
\end{equation}
\end{lemma}

\begin{proof}
Recall that there holds
\[
Q = \oneep \left( \FSh + \left( \rhoQ + \nu \right)\star |\: . \:| ^{-1} \right) - \oneep \left( \FSh \right)
\]
when $D(\nu,\nu)$ is small enough. We write this as 
$Q = \Pi_2-\Pi_1$
for short, with obvious definitions. Then
$ Q ^2 = \Pi_2 + \Pi_1 - \Pi_1 \Pi_2 - \Pi_2\Pi_1$
using $\Pi_i ^2 = \Pi_i$. But in this notation we have 
$\Qmm = \Pi_1 \Pi_2 \Pi_1 -\Pi_1$, and $\Qpp = \Pi_2+\Pi_1\Pi_2\Pi_1 - \Pi_2 \Pi_1-\Pi_1 \Pi_2$ which yield \eqref{formule Q}. Then \eqref{astuce energie 1} is obvious from the definition of the kinetic energy term 
$\Tro \left( \left(\FSh - \FSl \right) Q \right) = \Tr \left( \left| \FSh-\FSl \right| \left( \Qpp -\Qmm \right)\right)$. 
\end{proof}

The second step is less easy. It uses the decomposition $Q=Q_1+R_2$ and the special form of $Q_1$. As we explain below, the result should be true in a more general framework, but we present its proof in our context only, using the Bloch decomposition of $H^0_{\rm per}$ and the fact that the spectrum of this operator has a gap as crucial ingredients.

\begin{lemma}[\textbf{Leading order of the kinetic energy}]\label{lem: calcul sup 2}\mbox{}\\
There holds
\begin{equation}\label{calcul sup 2}
\Tr \left( \left| \FSh - \FSl \right| Q_1 ^2\right) = -\frac{1}{2}\Tr_0 (Q_1 V)
\end{equation}
where $V$ and $Q_1$ are defined in \eqref{Vm}.
\end{lemma}

To make the result plausible, let us present first a calculation which is valid for matrices. Let $C$ be an hermitian matrix and let us denote 
\[
\Pi_C = \one_{(-\infty,0)} (C).
\]
Writing for any operator $Q$
\[
\Tr (C Q)=  \Tr \left( |C| \left( \left(1-\Pi_C\right)Q\left(1-\Pi_C\right) - \Pi_C Q \Pi_C \right) \right)
\]
it follows that 
\begin{equation}\label{eq:formal}
\Tr\left( C (\Pi-\Pi_C)\right) = \Tr \left(|C| (\Pi-\Pi_C) ^2 \right),
\end{equation}
for any operator $C$ and projector $\Pi$. The proof goes along the very same lines as that of Lemma \ref{lem: calcul sup 1}, with the fact that $\Pi = \Pi ^2$ as only input.

Let us  now consider two operators $A,B$ and denote
\[
Q = \Pi_{A+ m B} - \Pi_A.
\] 
We will assume that $m\in \R$ is a small parameter and that $Q$ can be expanded as 
\[
Q = m \tilde{Q} + O (m ^2).
\] 
In the context of Lemma \ref{lem: calcul sup 2}, 
\[
A = \FSh - \FSl,\quad m = D(\nu,\nu) ^{1/2},\quad B = m ^{-1} V,\quad \tilde{Q} = Q_1/m
\]
with $V$ defined in \eqref{Vm} and $Q_1$ in \eqref{Q1_R2}.
Using \eqref{eq:formal} with $C= A + m B$ and $\Pi = \Pi_A$ we first compute
\[
\Tr \left( (A + m B) Q \right) = - \Tr \left( |A+ m B| Q ^2 \right).
\]
Inserting the expansion of $Q$ we thus have (at least formally)
\begin{equation}
\Tr \left( (A + m B) Q \right) = - \Tr \left( |A+ m B| Q ^2 \right) = - m ^2 \Tr (|A| \tilde{Q} ^{2}) + O(m ^3)\label{eq:formal 2}.
\end{equation}
On the other hand, by linearity and using \eqref{eq:formal} with $C = A$ and $\Pi = \Pi_{A+mB}$ we have
\begin{eqnarray}
\Tr \left( (A + m B) Q \right) &=& \Tr \left( A  Q \right) + m \Tr \left( B Q \right) \nonumber \\
&=& \Tr \left( |A| Q ^2 \right) + m \Tr \left( B Q \right)\nonumber \\
&=& m ^2 \Tr \left( |A| \tilde{Q} ^2 \right) + m ^2 \Tr \left( B \tilde{Q} \right)+O(m ^3) \label{eq:formal 1}.
\end{eqnarray}
Comparing \eqref{eq:formal 2} and \eqref{eq:formal 1} we obtain, to leading order in $m$
\[
\Tr \left( |A| \tilde{Q} ^2 \right) = -\frac{1}{2} \Tr \left( B \tilde{Q} \right)
\]
which is \eqref{calcul sup 2}.

It is elementary to justify the previous calculations for finite matrices, when $0$ is not in the spectrum of $A$, that is when $A$ has a gap in its spectrum~\cite{Kato}. In the infinite dimensional setting of this paper, the operators are not bounded and the trace $\Tr$ has to be replaced by the generalized trace $\Tro$. This generates serious difficulties if one tried to put the above sketch on a rigorous basis. For this reason we do not follow the simple strategy presented above and provide instead a proof of Lemma \ref{lem: calcul sup 2} relying on computations in a Bloch basis diagonalizing $A=\FSh-\FSl$..
\begin{proof}
We use the Bloch-wave decomposition of $\FSh$ \cite{ReeSim4}:
\begin{equation}\label{FSBloch1}
( \FSh f) (x) = \fintG \left( (\FSh)_q f_q \right)e^{iqx} dq.
\end{equation}
with
\begin{equation}\label{FSBloch2}
( \FSh )_q (x) = \sum_{n=1} ^{+\infty} \lam_{n,q} \ketl a_{n,q} \ketr \bral a_{n,q} \brar.
\end{equation}
Then we have (recall \eqref{eq:rHF-periodic})
\begin{eqnarray}
( \FSm )_q (x) &=& \sum_{n=1} ^{+\infty} \oneep (\lam_{n,q}) \ketl a_{n,q} \ketr \bral a_{n,q} \brar\nonumber \\
(1-\FSm )_q (x) &=& \sum_{n=1} ^{+\infty} \oneepp (\lam_{n,q}) \ketl a_{n,q} \ketr \bral a_{n,q} \brar \label{Pi plus bloch}.
\end{eqnarray}
We remark that as $V$ is a multiplication by a potential there holds $[V]_{p,q} = (V)_{p-q}$
which allows to compute the components of the Bloch matrix of $Q_1$, starting from \eqref{Q1_R2} 
\begin{eqnarray}\label{calcul Q1}
[Q_1]_{p,q} &=& \ointC \frac{1}{z-(\FSh)_p}(V)_{p-q} \frac{1}{z-(\FSh)_q}\\
&=& \ointC \left( \sum_l \frac{1}{z-\lam_{l,p}} \ketl a_{l,p} \ketr \bral a_{l,p} \brar\right) \left( \sum_n \frac{1}{z-\lam_{n,q}} \ketl (V)_{p-q} a_{n,q} \ketr \bral a_{n,q} \brar \right) \nonumber \\
&=& \sum_{l,n} \left( \ointC \frac{1}{(z-\lam_{l,p})(z-\lam_{n,q})}\right) \bral a_{l,p}  \brar (V)_{p-q} \ketl a_{n,q}\ketr \ketl a_{l,p} \ketr \bral a_{n,q} \brar \nonumber \\
&=& \sum_{l,n} \oneepp (\lam_{l,p}) \oneep (\lam_{n,q}) \frac{1}{\lam_{n,q}- \lam_{l,p}} \bral a_{l,p}  \brar (V)_{p-q} \ketl a_{n,q}\ketr \ketl a_{l,p} \ketr \bral a_{n,q} \brar \nonumber \\
&& \qquad+\sum_{l,n} \oneep (\lam_{l,p}) \oneepp (\lam_{n,q}) \frac{1}{\lam_{n,q}- \lam_{l,p}} \bral a_{l,p}  \brar (V)_{p-q} \ketl a_{n,q}\ketr \ketl a_{l,p} \ketr \bral a_{n,q} \brar \nonumber \\ 
&=& [C]_{p,q}+[D]_{p,q}.
\end{eqnarray}
We have used the residuum formula to evaluate the integral over the curve $\CC$. We then note that the operators $C$ and $D$ defined in the above 
formula satisfy $\Ran (C) \subset \Ker (C)$ and $\Ran (D) \subset \Ker(D)$, so $Q_1 ^2 = C D + D C$.

We compute the components of the Bloch matrix of $CD$ using the formula (the symbol $\fint = |\Gamma ^*| ^{-1} \int_{\Gamma ^*}$ denotes the mean value over the Brillouin zone $\Gamma ^*$)
\[
[CD]_{p,q} = \fint_{p'} C_{p,p'} D_{p',q} 
\]
and obtain
\begin{multline*}
[CD]_{p,q} = \sum_{l,n,n'} \oneepp (\lam_{l,p}) \oneep (\lam_{n,p'}) \frac{1}{(\lam_{n,p'}- \lam_{l,p})(\lam_{n,p'}- \lam_{n',q})}\\
\times \bral a_{l,p}  \brar (V)_{p-p'} \ketl a_{n,p'}\ketr
 \bral a_{n,p'}  \brar (V)_{p'-q} \ketl a_{n',q}\ketr \ketl a_{l,p} \ketr \bral a_{n',q} \brar.
\end{multline*}
Now 
$\left[ \left| \FSh- \FSl \right| CD \right]_{p,p} = \left| \FSh- \FSl \right|_p [CD]_{p,p}$
and thus, recalling \eqref{FSBloch1} and \eqref{FSBloch2},
\begin{eqnarray*}
\Tr \left( \left| \FSh- \FSl \right| CD \right) &=& \fint_p \Tr \left( \left| \FSh- \FSl \right|_p [CD]_{p,p} \right) \\
&=& \fint_p \fint_{p'} \sum_{l,n} \oneepp (\lam_{l,p}) \oneep (\lam_{n,p'}) \frac{|\lam_{l,p}-\FSl|}{\left( \lam_{n,p'} - \lam_{l,p} \right) ^2} \\
&& \qquad\qquad\times\left|\bral a_{l,p}  \brar (V)_{p-p'} \ketl a_{n,p'}\ketr \right| ^2
\end{eqnarray*}
where we have used that $\bral a_{l,p}  \brar (V)_{p-p'} \ketl a_{n,p'}\ketr = \overline{\bral a_{n,p'}  \brar (V)_{p'-p} \ketl a_{l,p}\ketr}$.
A similar computation leads to 
\begin{multline*}
\Tr \left( \left| \FSh- \FSl \right| DC \right) 
=\fint_p \fint_{p'} \sum_{l,n} \oneepp (\lam_{l,p}) \oneep (\lam_{n,p'}) \frac{|\lam_{n,p'}-\FSl|}{\left( \lam_{l,p} - \lam_{n,p'} \right) ^2} \\
\times \left|\bral a_{l,p}  \brar (V)_{p-p'} \ketl a_{n,p'}\ketr \right| ^2.
\end{multline*}
Summing the two contributions and noting that for $\lam_{l,p} > \FSl$ and $\lam_{n,p'} < \FSl$ we have
$|\lam_{l,p}-\FSl | + |\lam_{n,p'}-\FSl | = \lam_{l,p} - \lam_{n,p'}$,
we conclude that
\begin{multline}\label{fin tricky 1}
\Tr \left( \left| \FSh- \FSl \right| Q^2 \right) = \fint_p \fint_{p'} \sum_{l,n} \oneepp (\lam_{l,p}) \oneep (\lam_{n,p'}) \frac{1}{ \lam_{l,p} - \lam_{n,p'}} \\
\times \left|\bral a_{l,p}  \brar (V)_{p-p'} \ketl a_{n,p'}\ketr \right| ^2.
\end{multline}
An independent computation starting from \eqref{calcul Q1} yields
\begin{eqnarray*}
\Tr_0 (V Q) &=& \fint_p \fint_{p'} \sum_{l'} \bral a_{l',p} \brar (V)_{p-p'} (Q_1)_{p,p'} \ketl a_{l',p} \ketr\\
&=& \fint_p \fint_{p'} \sum_{l,n} \oneepp (\lam_{l,p'}) \oneep (\lam_{n,p}) \frac{1}{ \lam_{n,p} - \lam_{l,p'}}\left|\bral a_{l,p'}  \brar (V)_{p'-p} \ketl a_{n,p}\ketr \right| ^2 \\
&&\qquad+ \fint_p \fint_{p'} \sum_{l,n} \oneepp (\lam_{n,p}) \oneep (\lam_{l,p'}) \frac{1}{ \lam_{l,p'} - \lam_{n,p}} \left|\bral a_{l,p}  \brar (V)_{p-p'} \ketl a_{n,p'}\ketr \right| ^2\\
&=& - 2 \fint_p \fint_{p'} \sum_{l,n} \oneepp (\lam_{l,p}) \oneep (\lam_{n,p'}) \frac{1}{ \lam_{l,p} - \lam_{n,p'}} \left|\bral a_{l,p}  \brar (V)_{p-p'} \ketl a_{n,p'}\ketr \right| ^2
\end{eqnarray*}
and proves the lemma.
\end{proof}

We are now ready for the

\smallskip

\noindent \textbf{Proof of Proposition \ref{pro:to intermediary}.}
With Lemmas \ref{lem: calcul sup 1} and \ref{lem: calcul sup 2} at hand it essentially remains to bound the contribution of $R_2$ to the kinetic energy. This is done by using the estimates
\[
\Tro \left( \left| \FSh - \FSl \right| AB \right) \leq C \Vert A \Vert_{\Q} \Vert B \Vert_{\Q}.
\]
and 
\[
\left|\Tr_0(U A)\right| \leq C \left\Vert A \right\Vert_{Q} \left\Vert U \right\Vert_{L^2 + \coul '} 
\]
that hold true for any operators $A,B$ and any potential $U(x)$, see~\cite{CanDelLew-08a}. Using these in combination with \eqref{coulomb test} and 
\eqref{reste test} leads to
\begin{eqnarray*}
\left|\Tr \left( |\FSh - \FSl| Q ^2 \right) - \Tr \left( |\FSh - \FSl| Q_1 ^2 \right)\right| &\leq& C D(\nu,\nu) ^{3/2}\\
\left|\Tro (V Q ) - \Tro (V Q_1) \right| &\leq& C D(\nu,\nu) ^{3/2}.  
\end{eqnarray*}
Using that by definition of $V$, $\Tro (V Q ) = D(\rho_{Q}, \rho_Q + \nu)$, and combining Lemmas \ref{lem: calcul sup 1} and \ref{lem: calcul sup 2}, we infer 
\begin{align}\label{hum}
\crysint[\nu]=\cryse [Q,\nu] &= -\frac{1}{2}D(\rho_Q,\rho_Q+\nu) + D(\nu,\rho_Q) + \frac{1}{2} D(\rho_{Q}, \rho_Q) + O\left(D(\nu,\nu) ^{3/2}\right)\nonumber\\
&= \frac{1}{2}D(\rho_Q,\nu) + O\left(D(\nu,\nu) ^{3/2}\right).
\end{align}
Now, because of \eqref{eq:1st-order}, we can replace $\rho_Q$ by $-\K\nu$ in \eqref{hum} at the expense of an other error controlled by $D(\nu,\nu) ^{3/2}$. We thus obtain
$$\crysint[\nu] = - \frac{1}{2}D(\nu, \K\nu) + O\left(D(\nu,\nu) ^{3/2}\right),$$
as was claimed. \hfill\qed

\subsection{Towards the effective interaction $\ptgint$: two lemmas}\label{sec:macro 2}

Coming back to our main question, we have by Proposition~\ref{pro:to intermediary} and the estimate~\eqref{eq;bound_12_5_psi_m}
\begin{equation}\label{eq:clef crys int}
 m ^{-1} \crysint [m ^3 |\psi_m(m\cdot)| ^2] =  m ^{-1} \auxint [m ^3 |\psi_m(m\cdot)| ^2] + O(m ^{1/2}), 
\end{equation}
for any bounded sequence $(\psi_m)$ in $H^s(\R^3)$ with $s>1/4$. It therefore remains to study the limit of the right side of \eqref{eq:clef crys int}.

For a \emph{fixed defect} $\nu\in\coul$, we deduce from~\eqref{eq:clef crys int} and~\eqref{limit Bm} that
\begin{align*}
\lim_{m\to0}m ^{-1} \crysint [m ^3 \nu(m\cdot)] &=\lim_{m\to0}\frac{1}{2} \left( \int_{\R^3} B_m (\nu) \nu - D(\nu,\nu) \right)\\
&= \frac{1}{2} \left( \int_{\R^3} W_{\nu} \nu - D(\nu,\nu) \right) = \ptgint [\nu].
\end{align*}
We are now interested in proving the same kind of result for an $m$-dependent sequence of the form $\nu_m=|\psi_m|^2$, which is of course much more complicated. This is indeed not true in general, without appropriate assumptions on $(\psi_m)$. 

For instance, only assuming that $(\nu_m)$ is bounded in $\coul$ is certainly not sufficient to get the limit. Consider the example 
\begin{equation}
\nu_m=m^{1/2}m^{-3}\nu(x/m)
\label{eq:counter-example} 
\end{equation}
for a fixed density $\nu$, where the scaling is chosen such as to have $D(\nu_m,\nu_m)=D(\nu,\nu)$. In this case we have by~\eqref{eq:bound_B_m}
$$\int_{\R^3} B_m (\nu_m) \nu_m=m^{-1}D(\mathcal{U}_m\nu_m,(1+\Ll)^{-1}\mathcal{U}_m\nu_m)=D(\nu,(1+\Ll)^{-1}\nu)$$
which is \emph{not} close to 
$$\int_{\R^3}W_{\nu_m}\nu_m=\int_{\R^3}W_{\nu}\,\nu.$$
In this example the sequence $\nu_m$ actually lives at the microscopic scale because of our chosen scaling and a macroscopic behavior cannot be expected.\\

In order to be able to approximate $m ^{-1} \crysint [m ^3 \nu_m(m\cdot)]$ by its macroscopic counterpart $m ^{-1} \ptgint [m ^3 \nu_m(m\cdot)]$, we have to make sure that $\nu_m$ stays spread over a macroscopic region when $m\to0$. One way to ensure this is to impose that $\nu_m$ is locally compact in $L^{6/5}(\R^3)$. For $\nu_m=|\psi_m|^2$, this follows when $\psi_m$ is bounded in $H^s(\R^3)$ for some $s>1/4$ :

\begin{proposition}[\textbf{From $\auxint$ to $\ptgint$ in the locally compact case}]\label{pro:to PTg}\mbox{}\\
Let $(\psi_m)$ be a bounded sequence in $H^s (\R ^3)$, $s>1/4$. We have
\begin{equation}\label{to PTg}
\lim_{m\to0} \left( m ^{-1} \auxint [m ^3 |\psi_m(m\cdot)| ^2] - \ptgint [|\psi_m| ^2] \right) = 0.
\end{equation}
\end{proposition}

Combining \eqref{eq:clef crys int} and \eqref{to PTg} concludes the proof of Theorem \ref{theo:main crys}.
In the context of our polaron model, natural a priori $H^1$ bounds will be satisfied by minimizing sequences,  so the result will be applied with $s=1$.\\

The proof of Proposition \ref{pro:to PTg} uses the local compactness by resorting to standard techniques of nonlinear analysis. We write $\psi_m$ as a sum of pieces converging strongly (up to translation) in $L^{12/5}(\R^3)$, and receding from each other, plus a rest that is small in $L^{12/5}(\R^3)$ (a method that is usually called a \emph{bubble decomposition} in the literature). In order to conclude, we then need to prove two things : 
\begin{enumerate}
\item that one can pass to the limit for each compact piece of mass
\item that the interaction energy between two such pieces is negligible.
\end{enumerate}
In this section we prove two lemmas going in this direction, deferring the bubble analysis in itself and the conclusion of the proof of Proposition \ref{pro:to PTg} to the next section. 

Our first lemma deals with the case when $\nu_m$ converges strongly to some $\nu$ in $\coul$, up to a translation.

\begin{lemma}[\textbf{Self-interaction for strongly convergent sequences}]\label{lem:selfinteract Bm}\mbox{}\\
Let $(\nu_m)$ be a sequence of functions such that
\begin{equation}\label{asum lem1}
\nu_ m \to \nu \mbox{ strongly in } \coul
\end{equation}
when $m\to 0$. Let $(x_m)\subset\R^3$ be any sequence of translations. There holds
\begin{equation}\label{lem1,selfint}
\lim_{m\to0}\int_{\R ^3} B_m \left(\nu_m (\cdot-x_m)\right) \,\nu_m (\cdot-x_m) = \int_{\R ^3} W_{\nu}\, \nu.
\end{equation}
\end{lemma}

\begin{proof}
Using the fact that $B_m:\coul\to\coul'$ is bounded uniformly in $m$ by~\eqref{eq:estim_norm_Bm}, we deduce from the strong convergence of $\nu_m$ 
\[
\int_{\R ^3} B_m (\nu_m (\cdot-x_m)) \nu_m (\cdot-x_m) = \int_{\R ^3} B_m (\nu (\cdot-x_m)) \nu (\cdot-x_m) + o(1).
\]
The lattice spacing of the crystal is of order $m$ in our scaling, thus it is always possible to write
\[
x_m = \tau_m + y_m
\]
where $\tau_m\in m\cL$ is a translation of the scaled lattice and $|y_m| \leq C m$. 
It is clear from the definition of $\Ll$ in \cite{CanDelLew-08a} that $\Ll$ commutes with the translations of the original (unscaled) lattice, and this 
implies by our definition \eqref{Bm} that $B_m$ commutes with the translations of the scaled lattice $m\cL$. Thus 
\[
\int_{\R ^3} B_m (\nu (\cdot-x_m)) \nu (\cdot-x_m) = \int_{\R ^3} B_m (\nu(\cdot-y_m)) \nu(\cdot-y_m).
\]
Note that $\nu(\cdot-y_m)\to\nu$ strongly in $\coul$, hence we arrive at
$$\int_{\R ^3} B_m (\nu_m (\cdot-x_m)) \nu_m (\cdot-x_m) = \int_{\R ^3} B_m (\nu) \nu  + o(1)$$
and there only remains to use the result \eqref{limit Bm} proved in \cite{CanLew-10} to conclude the proof.
\end{proof}

We now provide the basic result allowing to neglect the interaction of two pieces of mass receding from each other. Note 
that the Pekar interaction of two such pieces of mass is easy to estimate because of the explicit form of its kernel, but the auxiliary 
interaction $\auxint$ has no such simple form and the argument is more involved. We actually have to come back to the technique of \cite{CanLew-10} used in the proof of \eqref{limit Bm}. 

\begin{lemma}[\textbf{Interaction of two pieces of mass receding from each other}]\label{lem:interact Bm}\mbox{}\\
Let $(\nu_{m,1})$ and $(\nu_{m,2})$ be two sequences of functions such that
\begin{equation}\label{asum lem2}
\nu_{m,i} \to \nu_i \mbox{ strongly in } \coul,
\end{equation}
for $i=1,2$ when $m\to 0$. Let $(x_{m,1})$ and $(x_{m,2})$ be two sequences of translations in $\R^3$ such that
$|x_{m,1} - x_{m,2}| \rightarrow + \infty$.
There holds
\begin{equation}\label{lem2,int}
\lim_{m\to0}\int_{\R ^3} B_m (\nu_{m,1} (\cdot-x_{m,1})) \nu_{m,2} (. - x_{m,2}) = 0.
\end{equation}
\end{lemma}

\begin{proof}
As in the proof of the previous lemma, the uniform boundedness of $B_m$ and the strong convergence \eqref{asum lem2} yield
\[
\int_{\R ^3} B_m (\nu_{m,1} (\cdot-x_{m,1})) \nu_{m,2} (. - x_{m,2}) = \int_{\R ^3} B_m (\nu_{1} (\cdot-x_{m,1})) \nu_{2} (. - x_{m,2}) + o(1).
\]
For the rest of the proof, we re-use some arguments of Lemma 7 in \cite{CanLew-10}. We define for $i=1,2$
\begin{equation*}
g_{i,m} = v_{c} ^{1/2} \nu_i(\cdot-x_{m,i})\in L^2(\R^3)
\end{equation*}
where $v_c$ is the Coulomb operator: 
\[
v_c (\nu) = \nu \star |\: . \: | ^{-1}.
\]
By density of regular functions in Fourier space, we can assume that both $\widehat{\nu_1}$ and $\widehat{\nu_2}$ are in $C^\ii_c(\R^3\setminus\{0\})$. Following \cite{CanLew-10} Lemma 7 we then have, for $m$ small enough,
\begin{align*}
\int_{\R ^3} B_m (\nu_{1} (\cdot-x_{m,1})) \nu_{2} (. - x_{m,2}) &= \int_{\R^3} dk \left< \left( \tilde{\ep}  ^{-1}\right)_{mk} e_0,e_0\right> \overline{\widehat{g_{1,m}} (k)} \widehat{g_{2,m}} (k)\\
&= \int_{\R^3} dk \left< \left( \tilde{\ep}  ^{-1}\right)_{mk} e_0,e_0\right> \overline{\widehat{g_1} (k)} \widehat{g_2} (k) e^{i\left( x_{m,2} - x_{m,1} \right)\cdot k}, 
\end{align*}
with $\ep = v_c \left(1+ \Ll \right) v_c ^{-1}$, $\tilde{\ep} = v_c ^{-1/2} \ep\, v_c ^{1/2}$ and $g_i = v_c ^{1/2} \nu_i$. 
In the second line we have used that $v_c$ commutes with multiplications in the Fourier space (it is itself such a multiplication). 
The factor $\left< \left( \tilde{\ep}  ^{-1}\right)_{mk} e_0,e_0\right>$ is uniformly bounded and tends to $|k|^2/k^T\dem\,k$ a.e., as proved in \cite{CanLew-10}, Lemma 6. By the dominated convergence theorem and Riemann-Lebesgue lemma, we thus have
\begin{equation}\label{preuve lem2 bis}
\lim_{m\to0}\int_{\R ^3} B_m (\nu_{1} (\cdot-x_{m,1})) \nu_{2} (. - x_{m,2})
= \lim_{m\to0}\int_{\Gamma^*} dk \frac{\overline{\widehat{\nu_1}(k)}\widehat{\nu_2}(k)}{k^T\dem\,k} e^{i\left( x_{m,2} - x_{m,1} \right)\cdot k}=0
\end{equation}
because $e^{i\left( x_{m,2} - x_{m,1} \right)\cdot k}$ converges to $0$ weakly-$\ast$ in $L ^{\infty}$. This concludes the proof.
\end{proof}

\subsection{Completion of the proof of Theorem~\ref{theo:main crys}}\label{sec:macro 3}

The proof of Theorem~\ref{theo:main crys} uses some classical tools of nonlinear analysis~\cite{Lieb-83,Lions-84,Lions-84b,Lions-87}, which are for instance recalled in the appendix of~\cite{LenLew-11} whose notation will be used here. The starting point is the `highest local mass' of a sequence $\boldsymbol{u}=(u_m)$, following ideas of Lieb~\cite{Lieb-83}.
\begin{definition}[\textbf{Highest local mass of a sequence}]\label{def:mass}\mbox{}\\
\it Let $\boldsymbol{u}=(u_n)$ be a bounded sequence in $H^s(\R ^3)$. We define 
\begin{multline}\label{local mass}
{\rm M }(\mathbf{u})= \sup \Big\lbrace \int_{\R ^3} |u| ^2,\: \exists (x_k) \subset \R ^3 : u_{n_k} (\cdot-x_k) 
\rightharpoonup u\\
 \mbox{ weakly in } H^s (\R ^3)  \mbox{ for some subsequence } (n_k) \Big\rbrace .
\end{multline}
\end{definition}

As first noted by Lions \cite{Lions-84,Lions-84b}, the above quantity plays a central role when controlling subcritical $L^p$ norms of the sequence, see Lemma 8 in \cite{LenLew-11}.

\begin{lemma}[\textbf{Control of subcritical $L^p$ norms}]\label{lem:util masse}\mbox{}\\
There exists a constant $C$, depending only on $s$, such that 
\begin{equation}\label{control Lp}
 \limsup_{n\to \infty} \int_{\R ^3 } |u_n| ^{2+\tfrac{4s}3} \leq C\, {\rm M}(\mathbf{u}) ^{\tfrac{2s}3} \limsup_{n\to \infty} \Vert u_n \Vert_{H ^s (\R ^3)}^2.
\end{equation}
\end{lemma}

We now recall a classical result allowing to locate the ``compact pieces of mass receding from each other'' we were alluding to before. 
Such bubble decompositions are ubiquitous in the literature (see, e.g., \cite{Struwe-84,BreCor-85,Lions-87,Gerard-98}). The following statement is taken from \cite{LenLew-11}.
\begin{lemma}[\textbf{Bubble decomposition of a bounded sequence in $H^s (\R ^3)$}]\label{lem:bubble}\mbox{}\\
Let $\mathbf{u}=(u_n)_n$ be a bounded sequence in $H^s(\R ^3)$ with $1/4<s<3/2$. For any $\ep > 0$ and any sequence~$0\leq R_n \to +\infty$, there exists 
$J\in \N$, sequences of functions $\mathbf{u_j} = (u_{j,n})_n, \: j=1\ldots J$ and $\mathbf{U_J} = (U_{J,n})_n$, sequences of translations 
$\mathbf{x_j} = (x_{j,n})_n, \: j=1\ldots J$ such that, along some subsequence (we omit to change indices)
\begin{equation}\label{bubble}
u_n - \sum_{j=1} ^J u_{j,n} (\cdot-x_{j,n}) - U_{J,n} \to 0 \mbox{ strongly in } H ^s (\R ^3).
\end{equation}
Moreover we have 
\begin{enumerate}
\item $u_{j,n} \to u_j \neq 0$ weakly in $H ^ s(\R ^3)$ and strongly in $L ^p (\R ^3)$ for any $2\leq p < 6/(3-2s)$ 
\item $\supp (u_{j,n}) \subset B(0,R_n)$ for any $j=1\ldots J$ and any $n\in \N$
\item $|x_{j,n}-x_{k,n}| \geq 5 R_n$ for any $j\neq k$ and any $n\in \N$
\item $\supp (U_{J,n}) \subset \R ^3 \setminus \cup_{j=1} ^J B(x_{j,n},2 R_n)$
\item ${\rm M}(\mathbf{U_J})\leq \ep$.
\end{enumerate}
\end{lemma}

We now complete the proof of Theorem \ref{theo:main crys}. The only missing ingredient is the 

\smallskip

\noindent\textbf{Proof of Proposition~\ref{pro:to PTg}.}
We apply\footnote{Strictly speaking we first extract a subsequence so as to consider a discrete set of $m$'s.} Lemma \ref{lem:bubble} to the sequence $(\psi_m)$, and denote $\nu_m=|\psi_m|^2$. We thus have, along some subsequence
\begin{equation}\label{bubble nu}
\left\Vert \psi_m - \varphi_m \right\Vert_{H ^s (\R ^3)} \to 0,\quad\text{ with }\  
\varphi_m = \sum_{j=1} ^J \varphi_{j,m}(\cdot-x_{j,m}) + \Psi_{J,m}.
\end{equation} 
The functions $\varphi_{j,m},\: j=1\ldots J$, $\Psi_{J,m}$ and the translations $x_{j,m}$ satisfy Items (1) to (5) in Lemma \ref{lem:bubble} 
for some $\ep$, $R_m$. We denote $\varphi_{j,\infty}$ the limit of the sequence $\varphi_{j,m}$. We further note that, as the functions appearing in \eqref{bubble nu} all have disjoint supports, there holds
\begin{equation}\label{bubble nu 2}
\left|\varphi_m\right|  ^2 = \sum_{j=1} ^J \left|\varphi_{j,m} (\cdot-x_{j,m})\right|^2 + \left|\Psi_{J,m}\right| ^2. 
\end{equation}
Notice that the assumption $s>1/4$ implies that $12/5<6/(3-2s)$ hence we have $|\varphi_{j,m}|^2\to |\varphi_{j,\ii}|^2$ strongly in $\coul$, by the Hardy-Littlewood-Sobolev inequality \cite{LieLos-01}. Also, using~\eqref{control Lp} and an interpolation argument, we can control the $L^{12/5}$ norm of the rest $\Psi_{j,m}$ by 
\begin{equation}
\limsup_{n\to \infty} \norm{\Psi_{J,m}}_{L^{12/5}(\R^3)}\leq C\, {\rm M}(\boldsymbol{\Psi_J}) ^{\tfrac{\theta s}{3+2s}} \limsup_{n\to \infty} \Vert \Psi_{J,m} \Vert_{H ^s (\R ^3)}^{1-\tfrac{2\theta s}{3+2s}}\leq C\,\varepsilon^{\tfrac{\theta s}{3+2s}}
\label{eq:estim-reste} 
\end{equation}
where $\theta>0$ is such that $5/12=3\theta/(6+4s)+(1-\theta)/q$ with $q=2$ for $3/10\leq s<3/2$ and $q=6/(3-2s)$ for $1/4<s<3/10$. The number $\varepsilon$ is that appearing in Item (5) of Lemma \ref{lem:bubble}.

We now claim that (along the appropriate subsequence)
\begin{equation}\label{split energy +}
\left|\int_{\R ^3} B_m (|\psi_m| ^2) |\psi_m| ^2 - \sum_{j = 1} ^J \int_{\R^3} W_{\left|\varphi_{j,\infty}\right| ^2} \left|\varphi_{j,\infty}\right| ^2 \right| \leq C \,\ep ^{\tfrac{2\theta s}{3+2s}} + o(1)_{m\to0}.
\end{equation}
To this end we use the decomposition \eqref{bubble nu} and \eqref{bubble nu 2} to obtain
\begin{eqnarray}\label{split energy}
\int_{\R ^3} B_m (|\psi_m| ^2) |\psi_m| ^2 &=& \sum_{j = 1} ^J  \int_{\R ^3} B_m \left(|\varphi_{j,m}(\cdot-x_{j,m})| ^2\right)|\varphi_{j,m}(\cdot-x_{j,m})| ^2 + \int_{\R ^3} B_m \left(|\Psi_{J,m}| ^2\right)|\Psi_{J,m}| ^2 \nonumber \\
&+& \sum_{j\neq k } \int_{\R ^3} B_m \left(|\varphi_{j,m}(\cdot-x_{j,m})| ^2\right)|\varphi_{k,m}(\cdot-x_{k,m})| ^2  \nonumber \\
&+& \sum_{j=1} ^J  \int_{\R ^3} B_m \left(|\varphi_{j,m}(\cdot-x_{j,m})| ^2\right)|\Psi_{J,m}| ^2 + o(1).
\end{eqnarray}
Now, using the continuity of $B_m$ from $\coul$ to $\coul'$ and the Hardy-Littlewood-Sobolev inequality, we obtain from~\eqref{eq:estim-reste}
\[
\int_{\R ^3} B_m \left(|\Psi_{J,m}| ^2\right)|\Psi_{J,m}| ^2 \leq C D(|\Psi_{J,m}| ^2,|\Psi_{J,m}| ^2) \leq C \left\Vert \Psi_{J,m} \right\Vert_{L ^{12/5}(\R^3)} ^4 \leq C \ep ^{\tfrac{4\theta s}{3+2s}},
\]
Similarly, 
\begin{multline}
\sum_{j=1} ^J\int_{\R ^3} B_m \left(|\varphi_{j,m}(\cdot-x_{j,m})| ^2\right)|\Psi_{J,m}| ^2\\
 \leq C \left\Vert \sum_{j=1} ^J|\varphi_{j,m}(\cdot-x_{j,m})|^2 \right\Vert_{L ^{6/5}(\R^3)} \left\Vert \Psi_{J,m} \right\Vert_{L ^{12/5}(\R^3)}^2 \leq C \ep^{\tfrac{2\theta s}{3+2s}}.
\end{multline}
In the second line we have used \eqref{bubble nu 2} to infer that
$$\left\Vert \sum_{j=1} ^J|\varphi_{j,m}(\cdot-x_{j,m})|^2 \right\Vert_{L ^{6/5}(\R^3)}\leq \left\Vert \,|\varphi_m|^2 \right\Vert_{L ^{6/5}(\R^3)}\leq C,$$
independently of $J$. On the other hand  
\[
\lim_{m\to0} \int_{\R ^3} B_m \left(|\varphi_{j,m}(\cdot-x_{j,m})| ^2\right)|\varphi_{j,m}(\cdot-x_{j,m})| ^2 = \int_{\R^3} W_{\left|\varphi_{j,\infty}\right| ^2} \left|\varphi_{j,\infty}\right| ^2
\]
and, for $j\neq k$,
$$
\lim_{m\to0} \int_{\R ^3} B_m \left(|\varphi_{j,m}(\cdot-x_{j,m})| ^2\right)|\varphi_{k,m}(\cdot-x_{k,m})| ^2 = 0
$$
by a direct application of Lemmas \ref{lem:selfinteract Bm} and \ref{lem:interact Bm}, respectively. We have thus proved the claim \eqref{split energy +}. 

A similar (but simpler) argument shows also that 
\begin{equation}\label{split energy ++}
\left|\int_{\R ^3} W _{|\psi_m| ^2} |\psi_m| ^2 - \sum_{j = 1} ^J \int_{\R^3} W_{\left|\varphi_{j,\infty}\right| ^2} \left|\varphi_{j,\infty}\right| ^2 \right| \leq C \,\ep ^{\tfrac{2\theta s}{3+2s}} + o(1)_{m\to0}.
\end{equation}
Indeed, one can decompose $\int_{\R ^3} W _{|\psi_m| ^2} |\psi_m| ^2$ like in \eqref{split energy}, and for any $\nu,\mu$ we have
\[
\left|\int_{\R ^3} W _{\nu} \mu\right| = \left|\frac{1}{4\pi} \int_{\R ^3} \frac{1}{k^T \dem k} \hat{\nu} (k) \hat{\mu} (k)\right| \leq C  \int_{\R ^3} \frac{1}{|k| ^2} |\hat{\nu} (k)|\, |\hat{\mu} (k)| \leq C \norm{\nu}_{L^{6/5}(\R^3)} \norm{\mu}_{L^{6/5}(\R^3)}.
\]
All the terms can then be dealt with exactly as before.

As a conclusion, comparing \eqref{split energy +} and \eqref{split energy ++}, we have shown that
$$\limsup_{m\to0}\left|\int_{\R ^3} B_m (|\psi_m| ^2) |\psi_m| ^2-\int_{\R ^3} W _{|\psi_m| ^2} |\psi_m| ^2\right|\leq C \,\ep ^{\tfrac{2\theta s}{3+2s}}.$$
Since $\ep$ can be chosen as small as we want, this proves the statement (for a convenient subsequence, but standard arguments yield the result along the whole sequence).\hfill\qed

\section{Proof of Theorem \ref{theo:main}}\label{sec:completion}

We have now all the tools needed for proving our main result.
Let $(\Psi_m)$ be a sequence of approximate minimizers in the sense of Definition \ref{def:minimiseur}. We write 
\begin{equation}\label{eq:split function}
\Psi_m = \EperM (\cdot/m) \polM
\end{equation}
and use the energy decoupling of Lemma \ref{lem:decouple} to obtain
\[
\E_m [\Psi_m] = N m^{-1} \Epere + \htotf [\polM].
\]
As a consequence of the $L^{\infty}$ estimate \eqref{estim uper} we deduce 
\begin{equation}\label{eq:split energy}
\E_m [\Psi_m] = N m^{-1} \Epere + \polf[\polM] (1+O(m)).
\end{equation}
with 
\begin{equation}\label{eq:functional polaron}
\polf [\Psi] := \left(\sum_{j=1} ^N \frac{1}{2} \int_{\R ^{3N}}|\nabla_j \Psi| ^2 +\sum_{1\leq k<\ell\leq N}\int_{\R ^{3N}}\frac{| \Psi|^2}{|x_k-x_\ell|}+ m ^{-1}\crysint \left[ m ^3 \rhoP (m\cdot) \right]\right).
\end{equation}
The meaning of \eqref{eq:split energy} is that the polaronic problem encoded in the functional \eqref{eq:functional polaron} has been completely decoupled from the microscopic oscillations due to $\FSp$.

\medskip

We now proceed to obtain a priori bounds on $\polM$ allowing to employ Theorem \ref{theo:main crys}. An upper bound to the energy is easily derived by taking a trial function of the form 
\begin{equation}\label{f test N}
\Ftest= \displaystyle \frac{\EperM (\cdot/m) \Psi }{\left\Vert \EperM(\cdot/m)  \Psi  \right\Vert_{L ^2 (\R^{3N})}}
\end{equation}
with $\Psi \in H ^1 (\R ^{3N})$ normalized in $L ^2 (\R^{3N})$ independent of $m$. Using the results of Section \ref{sec:periodic} as above we obtain 
\begin{equation}\label{eq:split energy sup}
\E_m [\Ftest] = N m^{-1} \Epere + \polf[\Psi] (1+O(m)).
\end{equation}
As $\Psi$ does not depend on $m$ we can use \eqref{eq:limit crys macro} to replace the complicated interaction $\crysint$ by the generalized Pekar interaction $\ptgint$
\[
\E_m [\Ftest] = N m^{-1} \Epere + \ptgf[\Psi] (1+O(m)) + o (1).
\]
As this holds for any $\Psi\in H^1 (\R ^{3N})$, we clearly have proved that 
\begin{equation}\label{eq:upper bound}
\limsup_{m\to0}E_m (N) \leq N \Eperelim + \ptge (N). 
\end{equation}

A rough lower bound to $\E_m[\Psi_m]$ is obtained by dropping the Coulomb term in \eqref{eq:split energy} and using Lemma \ref{lem:Fcrys} to bound the interaction with the crystal from below :
\begin{eqnarray}\label{eq:low bound rough}
\E_m[\Psi_m] &\geq& \frac{N\Epere}{m}+ \left(\sum_{j=1} ^N \frac{1}{2} \int_{\R ^{3N}}|\nabla_j \polM| ^2 \!- \frac{1}{2} D\left(m ^3\rho_{\polM}(m\cdot),m ^3\rho_{\polM}(m\cdot)\right)\right)\!\!(1+O(m)) \nonumber\\
&\geq& \frac{N\Epere}{m} + C_1 \left(\sum_{j=1} ^N \frac{1}{2} \int_{\R ^{3N}}|\nabla_j \polM| ^2 + \frac{1}{2} D\left(\rho_{\polM},\rho_{\polM}\right)\right)(1+O(m)) \nonumber \\ 
&&\qquad+ C_2 \left(\sum_{j=1} ^N \frac{1}{2} \int_{\R ^{3N}}|\nabla_j \polM| ^2 - \frac{1}{2} D\left(\rho_{\polM},\rho_{\polM}\right)\right)(1+O(m))
\end{eqnarray}
where $C_1$ and $C_2$ are two positive constants. The third term in the right-hand side of \eqref{eq:low bound rough} is given by the Pekar functional with a non optimal constant and is thus bounded below independently of $m$. Combining with the upper bound \eqref{eq:upper bound} we infer
\begin{equation}\label{eq:a priori}
\left(\sum_{j=1} ^N \frac{1}{2} \int_{\R ^{3N}}|\nabla_j \polM| ^2 + \frac{1}{2} D\left(\rho_{\polM},\rho_{\polM}\right)\right) \leq C.
\end{equation}
In the one-body case, this reduces to 
\[
\frac{1}{2} \int_{\R ^{3}}|\nabla \polM| ^2 + \frac{1}{2} D\left(|\polM| ^2, |\polM| ^2\right) \leq C,
\]
i.e. $\left(\polM\right)$ is a bounded sequence in $H ^1 (\R ^3)$. In the case $N\geq 2$, the Hoffmann-Ostenhof \cite{Hof-77} inequality
\[
\sum_{i=1} ^N \int_{\R ^{3N}} |\nabla_i \polM| ^2 dX \geq \int_{\R ^3}  \left|\nabla \sqrt{\rho_{\polM}}\right| ^2.
\]
allows to deduce from \eqref{eq:a priori} that $( \sqrt{\rho_{\polM}}) $ is a bounded sequence in $H^1 (\R ^3)$.

We can now apply Theorem \ref{theo:main crys} to the sequence $(\polM)$ when $N=1$, or $( \sqrt{\rho_{\polM}})$ when $N\geq2$, and obtain from \eqref{eq:split energy} and \eqref{eq:functional polaron} 
\[
\E_m [\Psi_m] = N m^{-1} \Epere + \ptgf [\polM] (1+O(m)) + o(1).
\]
On the other hand it is not difficult to see that 
\[
\left \Vert \polM \right\Vert_{L ^2 (\R ^{3N})} = 1 + O (m)
\]
so, writing 
\[
\tpolM = \left \Vert \polM \right\Vert_{L ^2 (\R ^{3N})}^{-1} \polM 
\]
we have
\begin{equation}\label{eq:borne inf finale}
\E_m [\Psi_m] = N m^{-1} \Epere + \ptgf \left[ \tpolM \right] (1+O(m)) + o(1).
\end{equation}
Recalling \eqref{eq:upper bound}, we deduce that $(\tpolM)$ is a minimizing sequence for $\ptge(N)$ and the conclusions of Theorem \ref{theo:main} follow by using Theorems \ref{theo:PTg1} and \ref{theo:PTgN}.\hfill\qed

%%%%%%%%%%%%%%%%%%%%%%%%%%%%%%%
%%%%%%%%%%%%%%%%%%%%%%%%%%%%%%%
% \bibliographystyle{siam}
% \bibliography{biblio}

\end{document}